\documentclass [11pt, letterpaper]{article}
\usepackage{fullpage}
\usepackage{amsmath}
\usepackage{amssymb}
\usepackage{graphicx}
\usepackage{mathrsfs}
\usepackage{wrapfig}
\usepackage{boxedminipage}
\usepackage{epsfig}
\usepackage{setspace}
\usepackage{subfigure}
\usepackage{float}
\usepackage{hyperref}
\usepackage{soul}
\usepackage{enumerate}
\newcommand{\reef}[1]{(\ref{#1})}
\begin{document}

\begin{flushright}
\phantom{{\tt arXiv:1008.3027}}
\end{flushright}

\bigskip
\bigskip
\bigskip

\begin{center} {\Large \bf Evolution of Holographic Entanglement Entropy }
  
  \bigskip

{\Large\bf  after }

\bigskip

{\Large\bf     Thermal and Electromagnetic Quenches}

\end{center}

\bigskip \bigskip \bigskip \bigskip

\centerline{\bf Tameem Albash, Clifford V. Johnson}

\bigskip
\bigskip
\bigskip

  \centerline{\it Department of Physics and Astronomy }
\centerline{\it University of
Southern California}
\centerline{\it Los Angeles, CA 90089-0484, U.S.A.}

\bigskip

\centerline{\small \tt talbash,  johnson1,  [at] usc.edu}

\bigskip
\bigskip


\begin{abstract} 
\noindent 
We study the evolution and scaling of the entanglement entropy after
two types of quenches for a 2+1 field theory, using a conjectured holographic
technique for its computation.  We study a thermal quench, dual to the addition of a
shell of uncharged matter to four dimensional Anti--de Sitter (AdS$_4$)
spacetime, and study the subsequent formation of a Schwarzschild black
hole.  We also study an electromagnetic quench, dual to the addition
of a shell of charged sources to AdS$_4$, following the subsequent
formation of an extremal dyonic black hole.  In these backgrounds we
consider the entanglement entropy of two types of geometries, the
infinite strip and the round disc, and find distinct behavior for
each. Some of our findings naturally supply results analogous to
observations made in the literature for lower dimensions, but we also
uncover several new phenomena, such as (in some cases) a discontinuity
in the time derivative of the entanglement entropy as it nears
saturation, and for the electromagnetic quench, a logarithmic growth
in the entanglement entropy with time for both the disc and strip,
before settling to saturation.  We briefly discuss the possible origin of the new phenomena in terms of the features of the conjectured dual field theory.

\end{abstract}
\newpage \baselineskip=18pt \setcounter{footnote}{0}


%
\section{Introduction}
%

The entanglement entropy of a quantum system has emerged as a useful
probe in a number of fields. For example, the condensed matter
community have recognized its utility as a diagnostic of phase
transitions, especially in contexts where an analogue of a
Landau--Ginsburg order parameter is either not accessible or
available.  The quantum information community employ it as a tool for
assessing such things as the computational complexity of a system (see
ref.\cite{Calabrese:2009pc} for a review). The quantum gravity
community have used it to address important issues such as the entropy
associated with a region of spacetime such as a black hole, where (for
example) the important relation between the entropy of an excised
region of spacetime (such as the interior of a black hole) and the
area of the spacetime surrounding it can be motivated in terms of
entanglement entropy (see {\it e.g.} ref.\cite{Srednicki:1993im}.)  In
such studies, it is very important to be able to extract information
about how the entanglement entropy scales with the size of a given
system, or evolves as a function of time.

Given a system, consider a region or subsystem which we can call
$\mathcal{A}$, with the remaining part of the system denoted by
$\mathcal{B}$. A definition of the entanglement entropy of
$\mathcal{A}$ with $\mathcal{B}$ is given by:
\begin{equation}
S_{\mathcal{A}} = - \mathrm{Tr}_{\mathcal{A}} \left( \rho_\mathcal{A} \ln \rho_\mathcal{A} \right)\ ,
\end{equation}
where $\rho_\mathcal{A}$ is the reduced density matrix of $\mathcal{A}$ given by tracing over the degrees of freedom of $\mathcal{B}$,
$\rho_\mathcal{A} = \mathrm{Tr}_{\mathcal{B}}( \rho) $,
%
where $\rho$ is the density matrix of the system.  When the system is in a pure state, \emph{i.e.,}
$\rho = \left| \Psi \rangle \langle \Psi \right|$, 
%
the entanglement entropy is a measure of the entanglement between the
degrees of freedom in $\mathcal{A}$ with those in $\mathcal{B}$.  If
the system starts in a pure state $| \Psi(0) \rangle$ and the evolution
of the system is unitary, then the system at subsequent times is also
a pure state $| \Psi(t) \rangle$, and so the evolution of the
entanglement entropy describes the evolution of quantum entanglement
in the system.

The evolution of entanglement entropy in a (1+1)--dimensional field
theory after a quench was studied in ref.~\cite{Calabrese:2005}.  The
system for $t<0$ is in the ground state (with a mass gap), and at $t =
0$, the magnetic field is set to the critical point of the
theory. Dividing the system into $\mathcal{A}$ (a line segment of
length $\ell$) and $\mathcal{B}$ (the rest), the analytical results
for the system showed that the entanglement entropy grows linearly
with time until it saturates at approximately $t = \ell/2$.
Ref.~\cite{Calabrese:2005} explained this as follows: at $t = 0$, the
quench results in excitations that propagate to the left and to the
right.  Since the system before the quench had a mass gap, only
excitations produced at around the same local region will be
entangled.  Contributions to the entanglement entropy occur when a
left/right mover remains in region $\mathcal{A}$ whereas the
right/left mover is in region $\mathcal{B}$.  So since at time $t$,
the left--right pair are separated by a distance $2 t$, saturation is
reached when excitations produced at the center of $\mathcal{A}$ reach
$\mathcal{B}$ which should be at $ 2 t \simeq \ell$.  It is of
interest to study more such systems and also to go beyond two
dimensions. It is hard to make progress for many interacting field
theories of interest, and so a search for powerful techniques from as
broad a range of sources seems prudent. One such source of tools is
the AdS/CFT
correspondence\cite{Maldacena:1997re,Witten:1998qj,Gubser:1998bc}.

A prescription for calculating the entanglement entropy
\cite{Ryu:2006ef} and its evolution \cite{Hubeny:2007xt} in the
context of AdS/CFT provides a new way to calculate the entanglement
entropy using geometrical techniques that are classical in spirit (for
a review see ref.\cite{Nishioka:2009un}). In an asymptotically Anti--de
Sitter (AdS) geometry, consider a slice at constant AdS radial
coordinate $ z = a$. Recall that this defines the dual field theory
(with one dimension fewer) as essentially residing on that slice in,
in the presence of a UV cutoff set by the position of the
slice. Sending the slice to the AdS boundary at infinity removes the
cutoff (see ref. \cite{Aharony:1999t} for a review).  On our $z=a$
slice, consider a region $\mathcal{A}$. Now find the minimal--area
surface $\gamma_{\mathcal{A}}$ bounded by the perimeter of
$\mathcal{A}$ and that extends  into the bulk of the geometry.
Then the entanglement entropy of region $\mathcal{A}$ with
$\mathcal{B}$ is given by:
\begin{equation}
S_{\mathcal{A}} = \frac{ \mathrm{Area}(\gamma_{\mathcal{A}})}{4 G_{\rm N}} \ ,
\end{equation}
where $G_{\rm N}$ is Newton's constant in the dual gravity
theory. This prescription for the entropy coincides nicely with
(1+1)--dimensional computations of the entanglement entropy, and has a
natural generalization to higher dimensional theories.  Note
  that there is no formal derivation of the prescription.  Steps have been
  made, such as in refs.~\cite{Fursaev:2006ih,Headrick:2010zt}, but
  they are not complete.  However, there is lots of evidence for the
  proposal. See \emph{e.g.,}
  refs.\cite{Solodukhin:2006xv,Hirata:2006jx,Nishioka:2006gr,Headrick:2007km,Solodukhin:2008dh,Casini:2008as}. A
  review of several of the issues can be found in
  ref.\cite{Nishioka:2009un}.  Further progress has been made recently in ref.~\cite{Casini:2011kv}.  Although there is no derivation available, in this paper we shall assume that this holographic prescription does give the correct result for the entanglement entropy in higher dimensions.  Our results are interpreted in line with this.  In particular, we find features that conform to expectations, and in addition we find new phenomena that we regard as predictions for the behavior of the field theory entanglement entropy.

Interesting recent papers, ref.\cite{Hubeny:2007xt,AbajoArrastia:2010yt},
presented a study of the time evolution of a (1+1)--dimensional system
after a thermal quench, by working in the AdS$_3$ geometry dual to
it. Since, \emph{via} the AdS/CFT correspondence, a thermal state in
the field theory is dual to a black hole on the gravity
side\cite{Witten:1998zw}, the authors started with AdS$_3$, perturbed
it with a shell of matter, and then followed the subsequent formation
of a black hole that took place, computing the entanglement entropy as
a function of time using the above geometric prescription. They used
the exact AdS$_3$ Vaidya metric to describe the formation process.

Their results confirmed a number of key features seen in the CFT
computation of ref.\cite{Calabrese:2005}. In particular, the linear
rise of the entanglement entropy with time, and the scaling of the
saturation time with the size of the system showing that the
effective propagation speed of the entanglement was the speed of
light. (Among the many interesting features of such computations is
the fact that the minimal surfaces involved must probe the region {\it
  behind} the apparent black hole horizon.)

The goal of the work we report on was primarily
two--fold. First, we wished to examine the question of higher
dimensions, and so we used an AdS$_4$ setting, relevant to the study
of dual 2+1 dimensional field theory. The question of how the entropy
evolves in time and scales with the size of the system can be examined,
and the physics of the growth of entanglement due to the propagation
of quanta ought to be more interesting since there are more shapes
available for region $\mathcal{A}$. We examine the physics of the
infinite strip and the round disc and indeed see rather interesting
physics to compare and contrast, again following an evolution after a
thermal quench modeled by a four dimensional Vaidya metric for
gravitational collapse. A second issue we wished to address was how to
do a different, non--thermal, type of quench in this AdS/CFT context,
and to compare our results for it to the thermal case. We constructed
a non--thermal quench by exciting electromagnetic sectors
instead. There is a natural $U(1)$ sector to which the physics couples
and so starting with pure AdS$_4$ we study and evolution of a shell
(using a charged generalization of the Vaidya metric) which ultimately
forms a zero temperature state at late times: an extremal black
hole. So the quench is non--thermal, and the excitation in our new
example is either by an external magnetic field or by a chemical
potential for charge density, or an admixture of the two that can be
freely dialed up using the four dimensional electric--magnetic duality
of the four dimensional gravity system.

In section~2 we give the general framework in which we are working,
setting out our notation and the general form of the computation. We
illustrate our methods by working out results for some familiar static
geometries, re--deriving old results and adding some new ones (those for
the case of extremal dyonic black holes). Section 3 presents our new
results for the thermal quench in higher dimensions, where we notice
several new features of the physics, including a kink in the time
evolution of the entropy as it heads to saturation, for large enough
system size. The case of the disc and the strip are thoroughly
examined and contrasted. Section 4 presents the electromagnetic
quench, discussing again many new features seen for the case of the
disc and the strip. We conclude in section~5, and there is a brief
appendix about the unitarity of the quench evolutions.

\section{General Framework}

In this section we present the setting for our computations of minimal
surfaces in asymptotically AdS$_4$ geometries. We will present some
examples of static test geometries in section 2.3, with some familiar
results and some new ones.  This will allow us to set up notation and
orient the discussion.

\subsection{Gravity Background}
%
We consider the Einstein--Maxwell action:
\begin{equation}
S = \frac{1}{16 \pi G_{\rm N}} \int d^4 x \sqrt{-G} \left( \mathcal{R} - 2 \Lambda - F^2 \right) + S_{ext} \ ,
\end{equation}
where the cosmological constant $\Lambda = -3 / R^2$ sets a length
scale $R$, and we will present the external sources we are interested
in shortly.  The equations of motion are given by:
\begin{equation}
\mathcal{R}_{\mu \nu} - \frac{1}{2} G_{\mu \nu} \left( \mathcal{R} - 2 \Lambda - F^2 \right) - 2 F_{\mu \lambda} F_{\nu}^{\phantom{j} \lambda} = 8 \pi G_4 T_{\mu \nu}^{\rm( ext)} \ ,
\end{equation}
\begin{equation}
\frac{1}{\sqrt{- G}} \partial_\mu \left( \sqrt{-G} F^{\mu \nu} \right) = 4 \pi G_4 J^{\nu}_{\rm (e-ext)}\ .
\end{equation}
We consider a solution to these equations of the form \cite{Chamorro:1994jg}:
\begin{eqnarray}
ds^2 &=& -V(r,v) dv^2 + 2 dr dv + \frac{r^2}{R^2} d \vec{x}^2  \ , \\
V(r,v) &=& \frac{r^2}{R^2} - \frac{2 m(v)}{r} + \frac{q_e(v)^2 + q_m(v)^2}{r^2} \ , \\
F_{r v} &=& \frac{q_e(v)}{r^2} \ , \quad F_{x y} = \frac{q_m(v)}{R^2} \ ,
\end{eqnarray}
which correspond to having sources given by:
\begin{equation}
4 \pi G_4 J^{\mu}_{\rm e-ext} = \frac{\dot{q}_e}{r^2} \delta^{\mu v}\ , \quad 8 \pi G_4 T_{\mu \nu}^{\rm ext} = - \frac{2}{r^3} \left( q_e \dot{q}_e + q_m \dot{q}_m - r \dot{m} \right) \delta_{\mu v} \delta_{\nu v} \ .
\end{equation}
This solution is a generalization of the Vaidya metric
\cite{Vaidya:1943pc,Vaidya:1951pc}.  Note that in terms of the Hodge
dual field, we can write:
\begin{equation}
\frac{1}{\sqrt{- G}} \partial_\mu \left( \sqrt{-G} (\ast F)^{\mu \nu} \right) = 4 \pi G_4 J^{\nu}_{\rm (m-ext)} \ , \quad 4 \pi G_4 J^\mu_{\rm (m - ext)} = \frac{\dot{q}_m}{r^2} \delta^{\mu v} \ .
\end{equation}
Also note that $(m(v), q_e(v), q_m(v))$ have dimensions of length.  It is convenient to work in terms of a coordinate $z$:
\begin{equation}
z = \frac{R^2}{r} \ ,
\end{equation}
such that we can write the metric as:
\begin{equation}
ds^2 = \frac{R^2}{z^2} \left(- f(z,v) dv^2  - 2 dz dv + d \vec{x}^2 \right) \ , \quad f(z,v) = 1 - \frac{2 m(v)}{R^4} z^3 + \frac{q_e(v)^2 + q_m(v)^2}{R^6} z^4 \ .
\end{equation}
The time coordinate $t$ of the dual field theory emerges (near the AdS boundary) as:
\begin{equation}
v \sim t - z \ .
\end{equation}
%
%
\subsection{Entanglement Entropy for the Strip and Disc}
%
\subsubsection{The Strip}
\begin{figure}[h]
\begin{center}
\subfigure[The strip.]{\includegraphics[width=3.0in]{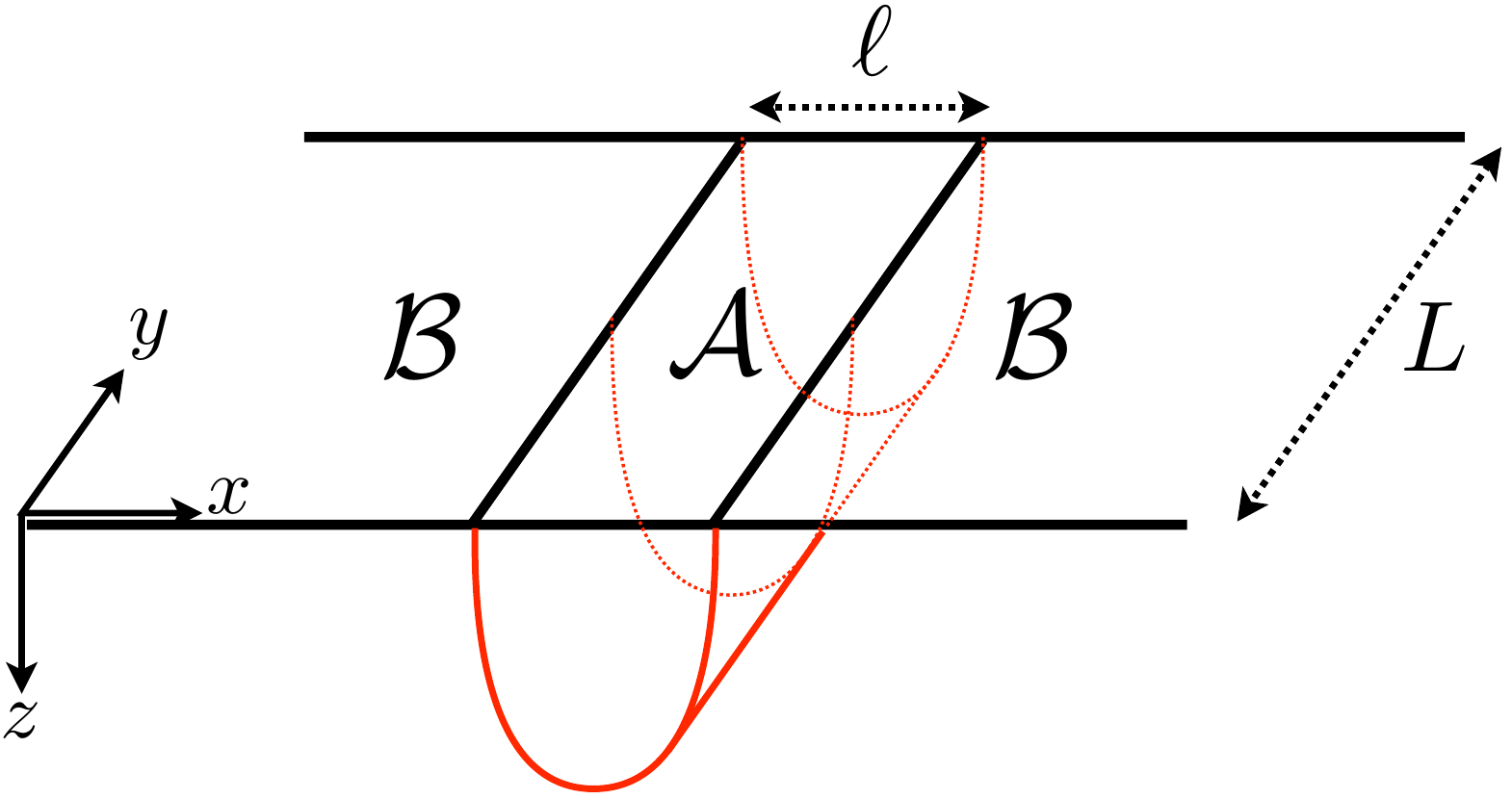}} \hspace{0.5cm}
\subfigure[The disc.]{\includegraphics[width=3.0in]{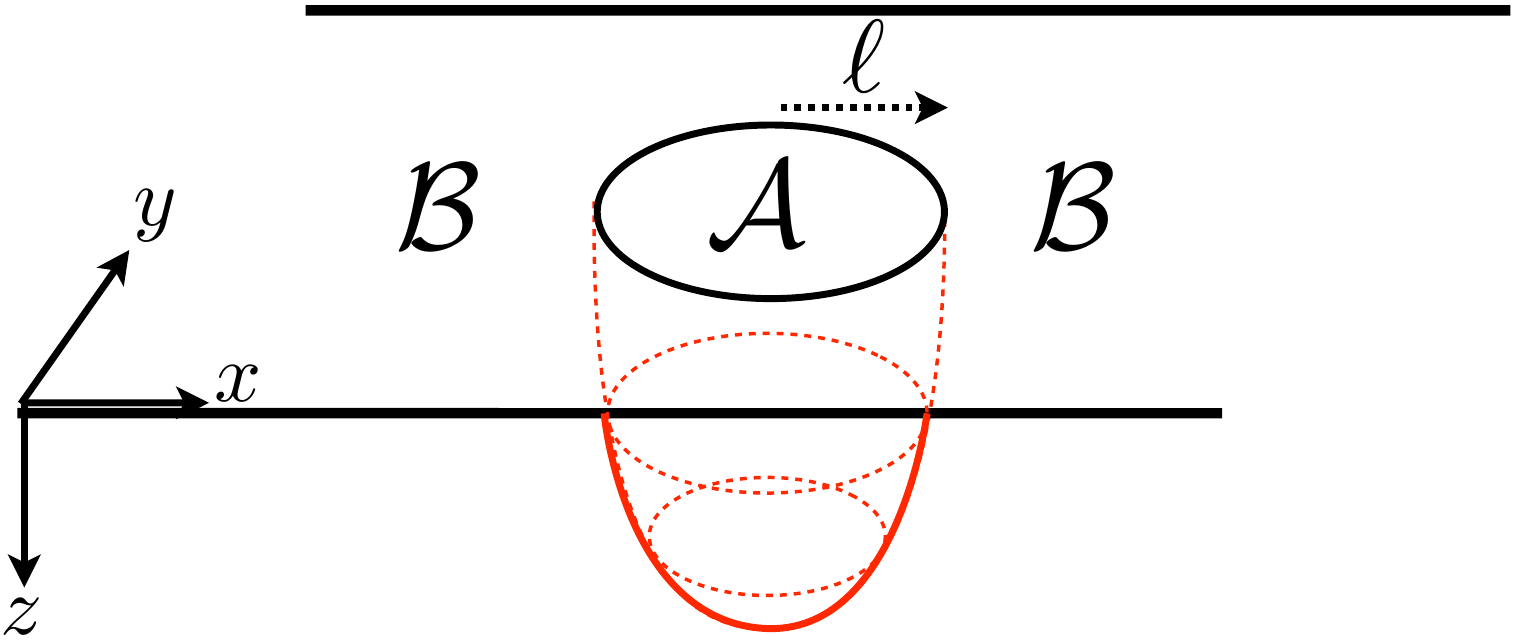}}
   \caption{\small Diagrams of the two shapes we will consider for region $\mathcal{A}$.}  \label{fig:setup_shapes}
   \end{center}
\end{figure}

%
We first consider calculating the entanglement entropy of an infinite
strip, which we denote as region $\mathcal{A}$ in the dual field
theory.  We consider the area of a two--dimensional
surface~$\gamma_{\mathcal{A}}$ that extends into the bulk that is
joined to the boundary of $\mathcal{A}$.  We depict this in figure
\ref{fig:setup_shapes}(a).

We parameterize the surface with coordinates:
\begin{equation}
\xi^1 = x \ , \quad \xi^2 = y \ , 
\end{equation}
At $ z = \epsilon$, very near the AdS boundary, ($\epsilon$ is the UV cutoff of the dual field theory), the region $\mathcal{A}$ will be chosen such that:
\begin{equation}
\mathcal{A} := x \in \left(- \frac{\ell}{2} , \frac{\ell}{2} \right) \ , \quad y \in \left( 0, L \right) \ .
\end{equation}
with $L$ taken to infinity.  The area of this surface is given by:
\begin{equation}
\mathrm{Area}(\gamma_\mathcal{A})  = L \int_{-\ell/2}^{\ell/2} dx \sqrt{h} \ ,
\end{equation}
where $h_{\alpha \beta}$ is the induced metric on the surface given by:
\begin{equation}
h_{\alpha \beta} = G_{\mu \nu} \frac{d x^{\mu}}{d \xi^{\alpha}} \frac{d x^{\nu}}{d \xi^{\beta}}  \ .
\end{equation}
By the symmetry of the strip, we consider an ansatz of the following form for the embedding profile of $\gamma_{\mathcal{A}}$:
\begin{equation}
v \equiv v(x) \ , \quad z \equiv z(x) \ , \quad z \left(\pm \ell/2 \right) = \epsilon \ , \quad v \left(\pm \ell/2 \right) = t - \epsilon \ ,
\end{equation}
such that the area is given by:
\begin{equation} \label{eqt:area_general}
\mathrm{Area}(\gamma_\mathcal{A})  = L R^2 \int_{-\ell/2}^{\ell/2} dx \frac{\sqrt{1 - f(v,z) v'(x)^2 - 2 v'(x) z'(x)} }{z(x)^2} \ .
\end{equation}
We denote the integrand in equation \reef{eqt:area_general} as the
Lagrangian $\mathcal{L}$ of our system.  Since there is no explicit
$x$ dependence, a constant of the motion is the Hamiltonian
$\mathcal{H}$ (if we think of the coordinate~$x$ as a time
coordinate),
\begin{equation} \label{eqt:hamiltonian}
\mathcal{H} = \mathcal{L} - v'(x) \frac{ \delta \mathcal{L}}{\delta v'(x)} - z'(x) \frac{ \delta \mathcal{L}}{\delta z'(x)} = \frac{1}{z(x)^2 \sqrt{1 - f(v,z) v'(x)^2 - 2 v'(x) z'(x)} } \ .
\end{equation}
By the symmetry of the problem, and by choice of origin of $x$, we
look for a turning point at $ x = 0$, and at this turning point, we
have:
\begin{equation}
v'(0) = z'(0) = 0 \ , \quad z(0) = z_\ast \ , \quad v(0) = v_\ast \ .
\end{equation}
Therefore, we have the result that $ \mathcal{H} = z_\ast^{-2}$.  Using this result and equation \reef{eqt:hamiltonian}, we can write the following conservation equation:
\begin{equation} \label{eqt:conservation}
1 - f(v,z) v'(x)^2 - 2 v'(x) z'(x) = \frac{z_\ast^4}{z(x)^4} \ .
\end{equation}
By taking the $x$--derivative of this equation, and substituting for $z''(x)$ in the two equations of motion, we get a single equation:
\begin{equation} \label{eqt:eom1}
-4 + 4 f(v,z) v'(x)^2 + 8 v'(x) z'(x) + 2 z(x) v''(x)  - z(x) v'(x)^2 \partial_z f(v,z) = 0 \ .
\end{equation}
Similarly, if we took the $x$--derivative of the conservation
equation, and substituting for $v''(x)$ in the two equations of
motion, we get a single equation:
\begin{equation} \label{eqt:eom2}\begin{array}{l}
 4 f(v,z)^2 v'(x)^2 - f(v,z) \left( 4 - 8 v'(x) z'(x) + z(x) v'(x)^2 \partial_z f(v,z) \right)  \\
 \quad \quad \quad \quad - z(x) \left( 2 z''(x) + v'(x) \left( 2 z'(x) \partial_z f(v,z) + v'(x) \partial_v f(v,z) \right) \right) = 0 \ . \\
 \end{array} 
\end{equation}
The ``on--shell'' area is now given simply by:
\begin{equation} \label{eqt:area}
\mathrm{Area}(\gamma_\mathcal{A}) = 2 L R^2 \int_{0}^{\ell/2} dx \frac{z_\ast^2}{z(x)^4} \ .
\end{equation}
%
\subsubsection{The Disc}
%

We may now consider the disc.  We consider a disc of radius $\ell$, and take the two dimensional surface $\gamma_\mathcal{A}$, which we depict in figure \ref{fig:setup_shapes}(b), to be parameterized by:
\begin{equation}
\xi^1 = r  \ , \quad \xi^2 = \phi \ .
\end{equation}
By the rotational symmetry of the problem, we can consider an ansatz of the form:
\begin{equation}
z \equiv z(r) \ , \quad v \equiv v(r) \ .
\end{equation}
This gives for the area:
\begin{equation} \label{eqt:area_disc}
\mathrm{Area}(\gamma_\mathcal{A}) = 2 \pi R^2 \int_0^{\ell} dr \frac{r}{z(r)^2} \sqrt{ 1 - f(v,z) v'(r)^2 - 2 v'(r) z'(r) } \ . 
\end{equation}
Because there is an explicit $r$ appearing in the Lagrangian, we do
not have a conserved quantity associated with the disc as we did for
the strip.  However, by the symmetry of the problem we again must
have:
\begin{equation}
z'(0) = v'(0) = 0 \ .
\end{equation}
The general equations of motion derived from the extremizing of the
area in equation \reef{eqt:area_disc} are straightforward to
derive. We do not write them here, in general, as they are rather
long. Later, specific cases will be written.
\subsection{Static Examples}
Following are computations for the entanglement entropy for three
static spacetime examples. In each case we present the results for the
strip and the disc.
\subsubsection{Pure AdS$_4$} \label{sec:pure_AdS4}
\begin{figure}[ht] %
   \centering
   \includegraphics[width=3in]{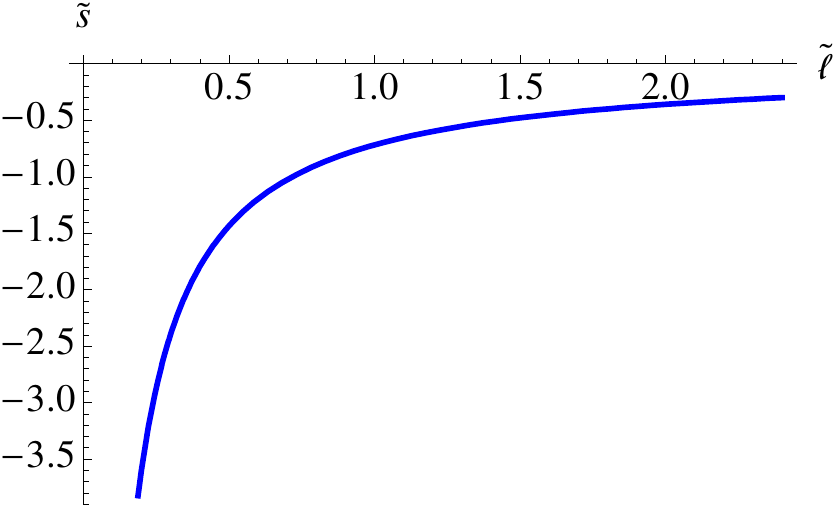} 
   \caption{\small The finite contribution to the entanglement entropy as a function of $\tilde{\ell} = \ell/R$ in the pure AdS$_4$ background.}
   \label{fig:AdS4}
\end{figure}

Let us calculate the entanglement entropy for pure AdS$_4$
\cite{Ryu:2006ef}.  This means taking~$f(v,z)=1$.  Although it is
simpler to exchange $v$ for $t$, let us proceed with the given choice
of coordinates.  A solution to the equations of motion \reef{eqt:eom1}
and \reef{eqt:eom2} is simply to take:
\begin{equation} \label{eqt:AdS4_ansatz}
v'(x) = - z'(x) \ .
\end{equation}
Then the conservation equation \reef{eqt:conservation} allows us to write:
\begin{equation} \label{eqt:AdS4_dzdx}
\frac{d z}{d x} =\pm \sqrt{ \frac{z_\ast^4}{z^4} - 1} \ ,
\end{equation}
where the plus sign is taken for $x < 0$ and the minus sign is taken for $x > 0$.  We can use this result to find a relationship between $z_\ast$ and $\ell$ by simply integrating equation~\reef{eqt:AdS4_dzdx}:
\begin{equation}
\frac{\ell}{2} = \int_\epsilon^{z_\ast} dz \ \left( \frac{z_\ast^4}{z^4} -1 \right)^{-1/2}  \ .
\end{equation}
Changing coordinates to $u = z^4/z_\ast^4$, we can write this integral in a more convenient form:
\begin{equation}
\frac{\ell}{2} =\frac{z_\ast}{4} \int_{\frac{\epsilon^4}{z_\ast^4}}^1 d u \ u^{-3/4} \left(1 - u \right)^{-1/2} = B\left( \frac{1}{4}, \frac{1}{2} \right) - B \left(\frac{\epsilon^4}{z_\ast^4};  \frac{1}{4}, \frac{1}{2} \right) \ ,
\end{equation}
where $B(a,b)$ is the Euler beta function, and $B(x; a,b)$ is the incomplete beta function.  In this expression, we can take $\epsilon \to 0$ without encountering any divergences, so our  result is simply:
\begin{equation}
\frac{\ell}{2} = z_\ast \frac{ \sqrt{\pi} \Gamma(3/4)}{\Gamma(1/4)} \ .
\end{equation}
Using the same procedures, we can write our ``on--shell'' area as:
\begin{equation}
\mathrm{Area}(\gamma_\mathcal{A}) = \frac{2 L R^2}{z_\ast^2} \int_a^{z_\ast} d z \ \frac{z_\ast^2}{z^2} \ \left( 1- \frac{z^4}{z_\ast^4} \right)^{-1/2} \ .
\end{equation}
Changing coordinates to $u = z^4/z_\ast^4$, we can write this integral in a convenient form:
\begin{equation}
\mathrm{Area}(\gamma_\mathcal{A})= \frac{2 L R^2}{4 z_\ast} \int_{\frac{\epsilon^4}{z_\ast^4}}^{1} d u \ u^{-5/4} \left( 1 - u \right)^{-1/2} = \frac{2 R L}{4 z_\ast} \left( B\left(- \frac{1}{4}, \frac{1}{2} \right) - B \left(\frac{\epsilon^4}{z_\ast^4}; - \frac{1}{4}, \frac{1}{2} \right) \right) \ .
\end{equation}
Expanding our expression for small $\epsilon$ (in order to capture the divergent contribution from our UV cutoff), we get our final result for the area \cite{Ryu:2006ef}:
\begin{equation} \label{eqt:AdS4_1}
\mathrm{Area}(\gamma_\mathcal{A})=  \frac{2 R^2 L}{\epsilon} + \frac{R^2}{2} \frac{ L}{z_\ast} \frac{\sqrt{\pi} \Gamma(-1/4)}{\Gamma(1/4)} = 2 R L \left( \frac{R}{\epsilon} + \frac{R}{2 \ell} \frac{\pi \Gamma(-1/4) \Gamma(3/4)}{\Gamma(1/4)^2} \right) \ .
\end{equation}
%
%
The divergent first term is proportional to the boundary of
$\mathcal{A}$.  We will be generally interested in the finite
contribution to the entropy given by the second term, which we denote
$s$:
%
%
\begin{equation} \label{eqt:finite_s_strip}
s =\frac{1}{4 G_{\rm N}} \left( \mathrm{Area}(\gamma_{\mathcal{A}}) - \frac{2 R^2 L}{\epsilon} \right)
\end{equation}
We will work in terms of the rescaled finite entropy $\tilde{s}$, a function of 
terms of the dimensionless length $\tilde{\ell} = \ell / R$,
\begin{equation} \label{eqt:tilde_s_strip}
\tilde{s} = \frac{4 G_{\rm N}}{2 R L} s  \ ,
\end{equation}
which we   plot in figure \ref{fig:AdS4}.

We can proceed in a similar fashion for the disc.  A solution to the equations of motion is to take the result of equation \reef{eqt:AdS4_ansatz}, which then gives us one equation for $z(r)$:
\begin{equation}
2 r \left( 1 + z'(r)^2 \right) + z(r) \left( z'(r) + z'(r)^3 + r z''(r) \right) = 0 \ .
\end{equation}
A solution to this equation is:
\begin{equation}
z(r)^2 + r^2 = \ell^2 \ , 
\end{equation}
which describes the extremal surface as a half--sphere (of radius $\ell$) in the bulk.  The area is given by:
\begin{equation}
\mathrm{Area}(\gamma_\mathcal{A}) = 2 \pi R^2 \ell \int_0^\ell dr \frac{r}{z(r)^3} \ .
\end{equation}
We use the fact that $ d z = - r dr  / z $ and $z_\ast = \ell$, and we change coordinates to $y = z / \ell$ to write the integral as:
\begin{equation}
\mathrm{Area}(\gamma_\mathcal{A})= 2 \pi R^2 \int_{\epsilon / \ell}^1 \frac{dy}{y^2} = 2 \pi R^2 \left( \frac{\ell}{\epsilon} - 1 \right) \ .
\end{equation}
If we define the finite contribution to the entropy by:
\begin{equation}  \label{eqt:finite_s_disc}
s = \frac{1}{4 G_{\rm N}} \left( \mathrm{Area}(\gamma_{\mathcal{A}}) - \frac{ 2 \pi R^2 \ell}{\epsilon} \right) \ ,
\end{equation}
and consider the rescaled entropy $\tilde{s}$
\begin{equation} \label{eqt:tilde_s_disc}
\tilde{s} \equiv \frac{4 G_{\rm N}}{ 2 \pi R^2} s \ ,
\end{equation}
then we find that $\tilde{s} = -1$. Note in particular that there is
no dependence on $\tilde{\ell}$, the size of the
disc\cite{Ryu:2006ef}.  We emphasize that this is a significant
difference from the strip and is a simple result of scale invariance
in the theory when a shape that preserves conformal invariance on the
boundary is taken for $\mathcal{A}$.
%
\subsubsection{AdS$_4$--Schwarzschild}
\begin{figure}[h] %
   \centering
   \includegraphics[width=3in]{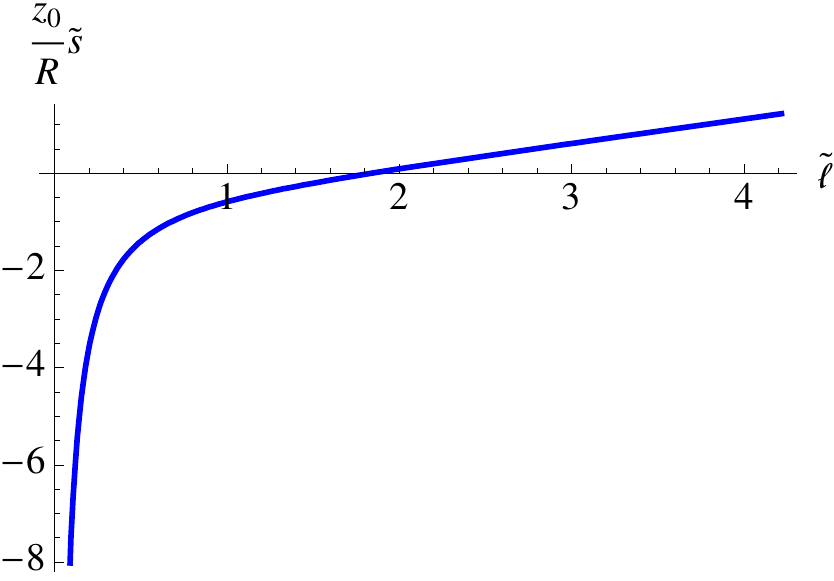} 
   \caption{\small The finite contribution to the entanglement entropy for the strip as a function of $\tilde{\ell} = \ell/z_0$ in the AdS$_4$--Schwarzschild background.}
   \label{fig:AdS4-Sch_strip}
\end{figure}

\begin{figure}[ht] %
   \centering
   \includegraphics[width=3in]{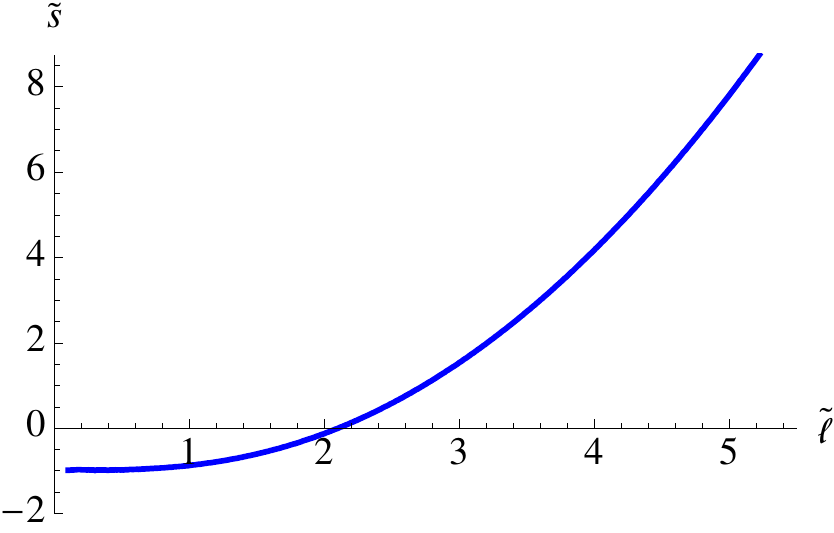} 
   \caption{\small The finite contribution to the entanglement entropy for the disc as a function of $\ell/z_0$ in the AdS$_4$--Schwarzschild background.}
   \label{fig:AdS4-Sch-Disc}
\end{figure}

%
To study the case of AdS$_4$--Schwarzschild\cite{Ryu:2006ef}, we take $f(v,z) = 1 - 2 m
z^3 / R^4$.  In this geometry, the event horizon is located at a
radial distance  $z_0$ given by:
\begin{equation}
z_0 = \left( \frac{2m}{ R^4} \right)^{-1/3} \ .
\end{equation}
We once again start with the strip.  A solution to the equations of motion is to take:
\begin{equation} \label{eqt:v}
v(x) = t + g(z(x))  \ , \quad \partial_{z(x)} g(z(x)) = - f(z(x))^{-1} \ .
\end{equation}
The remaining equation of motion is given by:
\begin{equation} \label{eqt:eom_nov}
- 2 z(x) z''(x) + z'(x)^2 \left( -4 + z(x) \frac{\partial_{z(x)} f(z(x))}{f(z(x))} \right) - 4 f(z) = 0 \ .
\end{equation}
The conservation equation is now given by:
\begin{equation} \label{eqt:cons_nov}
1 + \frac{z'(x)^2}{f(z(x))} = \frac{z_\ast^4}{z(x)^4} \ ,
\end{equation}
which allows us to write:
\begin{equation} \label{eqt:dz/dz_nov}
\frac{d z}{d x}  = \pm \sqrt{ f(z(x)) \left(  \frac{z_\ast^4}{z(x)^4 } - 1 \right)} \ .
\end{equation}
In turn, the area can be written as:
\begin{equation} \label{eqt:A_nov}
\mathrm{Area}(\gamma_{\mathcal{A}}) = \frac{2 L R^2}{z_\ast^2} \int_{\epsilon}^{z_\ast} dz \frac{z_\ast^2}{z^2} \left( f(z) \left( 1 - \frac{z^4}{z_\ast^4} \right) \right)^{-1/2} \ ,
\end{equation}
and the length is given by:
\begin{equation} \label{eqt:l_nov}
\frac{\ell}{2} = \int_0^{z_\ast} dz \frac{z^2}{z_\ast^2} \left( f(z) \left( 1 - \frac{z^4}{z_\ast^4} \right) \right)^{-1/2} \ .
\end{equation}
It is convenient at this point to define dimensionless variables:
\begin{equation} \label{eqt:dimensionless}
\tilde{z} = \frac{z}{z_0} \ , \quad \tilde{x}  = \frac{x}{z_0} \ ,
\end{equation}
such that:
\begin{equation}
\mathrm{Area}(\gamma_{\mathcal{A}})  = \frac{2 L R^2}{z_0} \int_{\tilde{\epsilon}}^{\tilde{z}_\ast} d \tilde{z} \tilde{z}^{-2} \left( \left(1-\tilde{z}^3\right) \left(1 - \frac{\tilde{z}^4}{\tilde{z}_\ast^4} \right) \right)^{-1/2} \ ,
\end{equation}
and
\begin{equation} \label{eqt:ell}
\frac{\tilde{\ell}}{2} = \int_0^{\tilde{z}_\ast} d \tilde{z} \frac{\tilde{z}^2}{\tilde{z}_\ast^2} \left( (1-\tilde{z}^3) \left( 1 - \frac{\tilde{z}^4}{\tilde{z}_\ast^4} \right) \right)^{-1/2} \ .
\end{equation}
Since there is no analytic expression for the area, we wish to express
the finite contribution to the entanglement in a way that is
consistent with our pure AdS$_4$ result using equation
\reef{eqt:finite_s_strip}.  Furthermore, we can define a rescaled
entropy $\tilde{s}$ as in equation \reef{eqt:tilde_s_strip}:
\begin{equation}
\tilde{s} =  \frac{R}{z_0} \left[   \int_{\tilde{\epsilon}}^{\tilde{z}_\ast} d \tilde{z} \tilde{z}^{-2} \left( \left(1-\tilde{z}^3\right) \left(1 - \frac{\tilde{z}^4}{\tilde{z}_\ast^4} \right) \right)^{-1/2} - \frac{1}{\tilde{\epsilon}} \right]
\end{equation}
 We present a plot of this finite contribution to the area in figure \ref{fig:AdS4-Sch_strip}.
In particular, note that for large
$\tilde{\ell}$, the result for $z_0 \tilde{s} / R$ scales as $\tilde{\ell}/2$ which is exactly what is needed to recover the
entropy--area law, \emph{i.e.}:
\begin{equation}
  (4 G_{\rm N}) s \sim \frac{R^2}{z_0^2} L \ell \ ,  \ \mathrm{for} \ \ell/z_0 \ \mathrm{large}\ .
\end{equation}
We can perform a similar computation for the disc.  The area can be written as:
\begin{equation}
\mathrm{Area}(\gamma_{\mathrm{A}}) = 2 \pi R^2 \int_0^{\tilde{\ell}} d \tilde{r} \frac{\tilde{r}}{\tilde{z}^2} \sqrt{ 1 + \frac{\tilde{z}'(\tilde{r})^2}{1 - \tilde{z}^3} }\ .
\end{equation}
We write the finite contribution to the entropy $s$ as in equation
\reef{eqt:finite_s_disc}, and we write the rescaled entropy
$\tilde{s}$ as in equation \reef{eqt:tilde_s_disc}:
\begin{equation}
\tilde{s} = \int_0^{\tilde{\ell}} d \tilde{r} \frac{\tilde{r}}{\tilde{z}^2} \sqrt{ 1 + \frac{\tilde{z}'(\tilde{r})^2}{1 - \tilde{z}^3} } - \frac{\tilde{\ell}}{\tilde{\epsilon}}\ . 
\end{equation}
We present the behavior of the area in figure \ref{fig:AdS4-Sch-Disc}.
We see that for small $\tilde{\ell}$, the contribution matches that of
pure AdS$_4$, as the surface remains close to the AdS boundary.  For
large $\tilde{\ell}$, the behavior of the area is quadratic in
$\tilde{\ell}$ as is to be expected, since most of the contribution
will come from the surface behaving as a disc at the event horizon.
Again we find that for large $\tilde{\ell}$, the entropy $\tilde{s}$
scales as $\tilde{\ell}^2 /2$, which again recovers the entropy--area
law:
\begin{equation}
(4 G_{\rm N})s \sim  \frac{ R^2}{z_0^2} \pi \ell^2 \ ,  \ \mathrm{for} \ \ell/z_0 \ \mathrm{large}\ .
\end{equation}
%
\subsubsection{Extremal AdS$_4$--Dyonic} \label{section:Dyonic}
\begin{figure}[h] %
   \centering
   \includegraphics[width=3in]{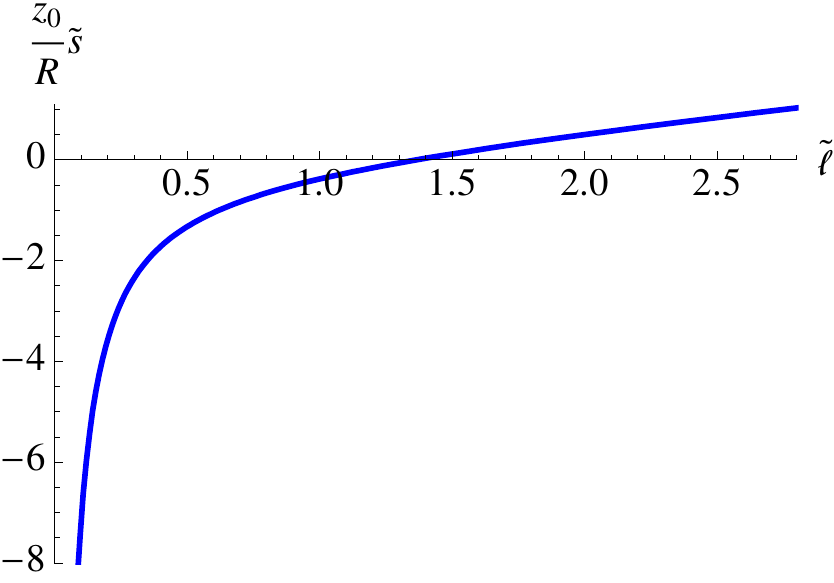} 
   \caption{\small The finite contribution to the entanglement entropy for the disc as a function of $\ell/R$ in the AdS$_4$ Dyonic black hole background.}
   \label{fig:AdS4-Dyon_strip}
\end{figure}

\begin{figure}[h] %
   \centering
   \includegraphics[width=3in]{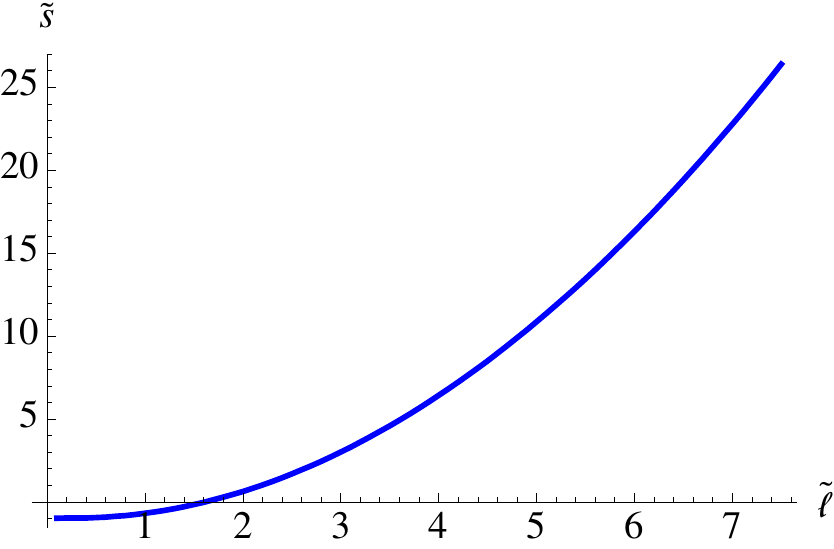} 
   \caption{\small The finite contribution to the entanglement entropy for the disc as a function of $\ell/R$ in the AdS$_4$ Dyonic black hole background.}
   \label{fig:AdS4-Dyon_disc}
\end{figure}

%
We now consider the extremal AdS$_4$ dyonic black hole. The
entanglement entropy for this static case has not been studied (as far
as we know) in the literature, and so the results are new. This is a
useful preparatory computation for the electromagnetic quench we will
perform in a later section. We have:
\begin{equation}
f(z) = 1 - 4 \left( \frac{z}{z_0} \right)^3 + 3 \left( \frac{z}{z_0} \right)^4 \ ,
\end{equation}
where we have used that:
\begin{equation}
Q^2 = q_e^2 + q_m^2 = \frac{3 m}{2 z_0} R^2 \ , \quad z_0^3 = \frac{2 R^4}{m} \ .
\end{equation}
Equations \reef{eqt:v}, \reef{eqt:eom_nov}, \reef{eqt:cons_nov},
\reef{eqt:dz/dz_nov}, \reef{eqt:A_nov}, and \reef{eqt:l_nov} remain
unchanged.  We can define the finite part of the entropy using
equation \reef{eqt:finite_s_strip} and the rescaled entropy using
equation \reef{eqt:tilde_s_strip}.  We present the finite contribution
to the area in figure \ref{fig:AdS4-Dyon_strip}.  As in the case of
the Schwarzschild black hole, for large~$\tilde{\ell}$ we recover the
entropy--area law with a linear dependence in $\tilde{\ell}$.
Performing a similar calculation for the disc gives the results in
figure \ref{fig:AdS4-Dyon_disc}.  We see again that for $\tilde{\ell}$
large, the behavior is such that we recover the entropy--area law, with a quadratic dependence in $\tilde{\ell}$.

\begin{figure}[h]
\begin{center}
\subfigure[Embedding with $z_\ast < z_0$]{\includegraphics[width=3.0in]{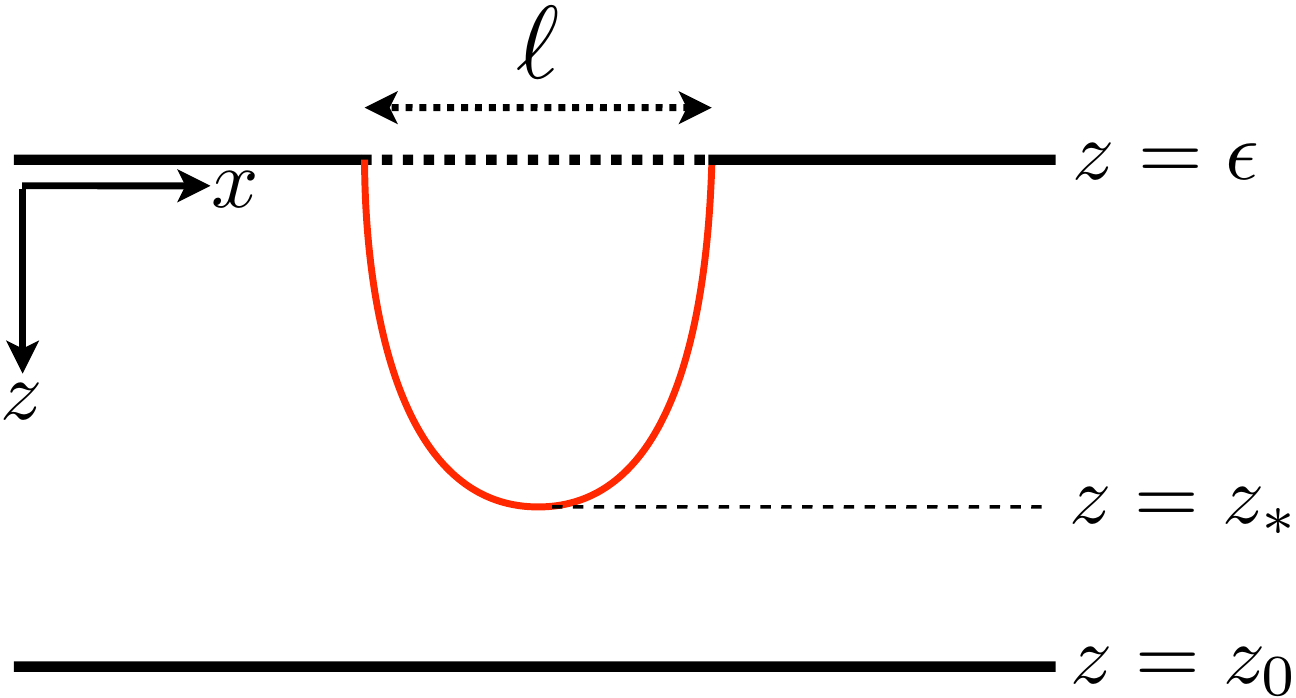}} \hspace{0.5cm}
\subfigure[Embedding with $z_\ast = z_0$]{\includegraphics[width=3.0in]{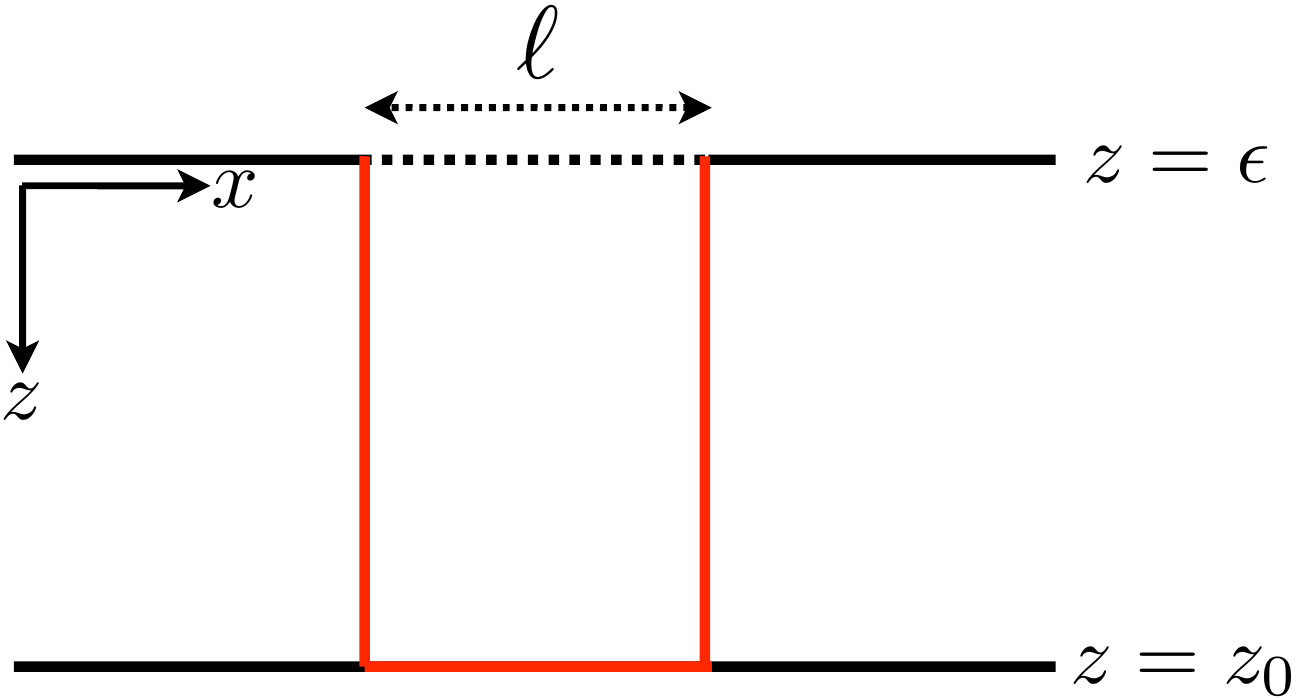}}
   \caption{\small Diagrams of the two types of embeddings in the presence of a black hole.}  \label{fig:embeddings}
   \end{center}
\end{figure}

A new feature of this background is that the area integral exhibits a
logarithmic divergence as $z_\star$ approaches the event horizon.  This
is a simple result of the fact that the event horizon has a double
zero. This feature will play a role in the evolution of our
electromagnetic quench discussed later.
To see this more explicitly, as an extreme example, consider the embedding with $z_\ast =
z_0$.  This embedding consists of a piece that extends along the event
horizon, and then two pieces that extend from the AdS boundary at $x =
\pm \ell/2$ (for the strip) to the event horizon.  We present a
depiction of this embedding in figure \ref{fig:embeddings}.
The contribution to the area of the pieces that fall towards the event horizon is given by:
\begin{equation}
\mathrm{Area} (\gamma_{\mathcal{A}}) = 2 L \int_{\epsilon}^{z_0} \frac{R^2}{z^2} f(z)^{-1/2} \ .
\end{equation}
Near the event horizon, we can write:
\begin{equation}
f(z \to z_0 ) = \frac{6}{z_0^2} \left( z_0 - z \right)^2 \ ,
\end{equation}
and so we have:
\begin{equation}
\mathrm{Area} (\gamma_{\mathcal{A}}) \sim \int^{z_0 }\frac{d z}{z_0 - z} \ ,
\end{equation}
which contributes a logarithmic divergence to the area.
%
\section{The Thermal Quench}
%
We begin by extending the very interesting work of
refs.~\cite{Hubeny:2007xt,AbajoArrastia:2010yt} to one dimension higher, performing a
thermal quench. We will get several new results, and contrast them
with several more new results that we obtain for a new,
electromagnetic, holographic quench in a later section.
The idea of the holographic quench studied in
refs.\cite{Hubeny:2007xt,AbajoArrastia:2010yt} (see also ref.\cite{Das:2010yw}) is
that the system is simply AdS at early times, and then at some later
time it is strongly perturbed by a shell of collapsing null dust,
which subsequently forms a Schwarzschild black hole at late times,
representing a thermal state in the holographically dual
theory\cite{Witten:1998zw}. The point of the exercise is to study the
evolution of the entanglement entropy over time after the
quench\footnote{This evolution is indeed unitary, as it ought to
  be~\cite{AbajoArrastia:2010yt}. See the appendix  for some clarifying
  comments on that issue. We thank E. Lopez for a discussion on this
  point.}.  Generically, it saturates after a characteristic time. In
1+1 dimensions the behaviour is rather simple, with a linear rise in
the entropy followed by a distinct leveling off to saturation, as
observed in a magnetically quenched system in
ref.\cite{Calabrese:2005}, and a thermally quenched system using
holographic duality to asymptotically AdS$_3$ geometries in
refs.\cite{Hubeny:2007xt,AbajoArrastia:2010yt}. The slope of the line reflects the
fact that the effects of the quench propagate at the speed of light in
the single dimension available. We shall see more complicated
possibilities in higher dimensions and in the next section when we do an
electromagnetic (non--thermal) quench.

\begin{figure}[h] %
   \centering
   \includegraphics[width=3in]{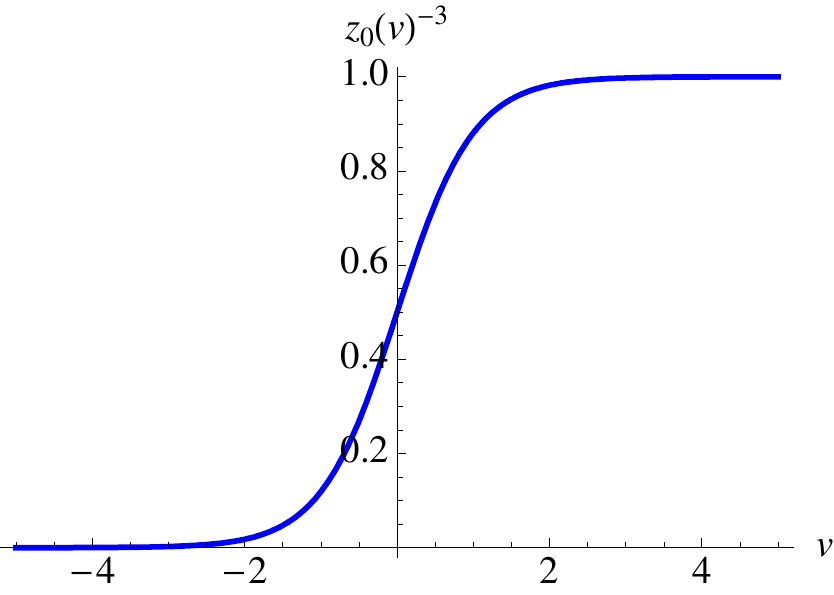} 
   \caption{\small The shape of the mass function $m(v)$,
     generating the quench from pure AdS$_4$ to a finite mass black
     hole.}
   \label{fig:tanhplot}
\end{figure}

To achieve the quench, we consider:
\begin{equation}
f(v,z) = 1 - \left(\frac{z}{z_0(v)} \right)^3 \ , \quad m(v) = z_0(v)^{-3} \ .
\end{equation}
and take the following functional form for $z_0(v)$:
\begin{equation} \label{eqt:z0}
z_0(v) = z_\infty \left( \frac{1+ \tanh(v/v_0)}{2} \right)^{-1/3} \ .
\end{equation}
This form for $z_0(v)$ is such that the background is pure AdS$_4$ in
the infinite past and an AdS$_4$--Schwarzschild black hole in the
infinite future. See figure~\ref{fig:tanhplot}.
The parameter $z_\infty$ can be thought of as being related to the
final temperature of the background at late times:
\begin{equation}
T = \frac{3}{4 \pi} z_\infty^{-1}\ ,
\end{equation}
and the parameter $v_0$ is related to the time over which the quench
occurs.  As in the case of the static AdS$_4$--Schwarzschild black
hole, we rescale $z$ and $x$ by $z_\infty$ to define dimensionless
coordinates~$\tilde{z}$ and~$\tilde{x}$.  We proceed to solve
equations \reef{eqt:eom1} and \reef{eqt:eom2} numerically using a shooting method with the following
initial conditions:
\begin{equation}
\tilde{v}'(0) = \tilde{z}'(0) = 0 \ , \quad \tilde{z}(0) = \tilde{z}_\ast \ , \quad \tilde{v}(0) = \tilde{v}_\ast \ .
\end{equation}
Near the AdS boundary (\emph{i.e.} when $\tilde{x} = \tilde{\ell}/2$ for the strip or when $\tilde{r} = \tilde{\ell}$ for the disc), we extract the following information:
\begin{equation}
\tilde{z}(\tilde{\ell}/2) = \tilde{\epsilon} \ , \quad \tilde{v}(\tilde{\ell}/2) = \tilde{t} - \tilde{\epsilon} \ ,
\end{equation}
Because of the form we choose for our quench, the quench starts at approximately $ \tilde{t} \approx - 2 \tilde{v}_0$. 
%
\subsection{The Strip} \label{sec:thermal_strip}
\begin{figure}[h] %
   \centering
   \includegraphics[width=3in]{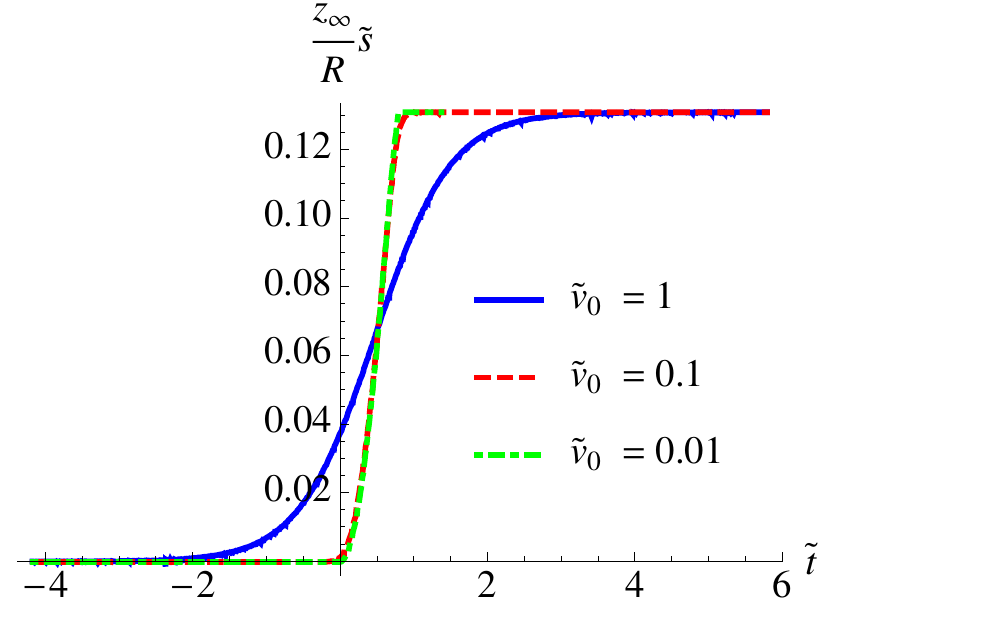} 
   \caption{\small The evolution of $\tilde{s}$ for different quench speeds for $\tilde{\ell}=1$.}
   \label{fig:thermal_vary_v0}
\end{figure}

%
After the rescaling, the area of the surface given in equation \reef{eqt:area} is given by:
\begin{equation} \label{eqt:quench_area_strip}
\mathrm{Area}(\gamma_\mathcal{A}) = \frac{2 L R^2}{z_\infty} \int_0^{\tilde{\ell}/2} d \tilde{x} \frac{\tilde{z}_\ast^2}{\tilde{z}(\tilde{x})^4} \ .
\end{equation}
We define the finite contribution to the entanglement entropy $s$ as
in equation \reef{eqt:finite_s_strip}, but we modify our definition of
the rescaled entropy from equation \reef{eqt:tilde_s_strip} by
subtracting the pure AdS$_4$ result we have found in section
\ref{sec:pure_AdS4}:
\begin{equation} \label{eqt:quench_s_strip} \tilde{s} =
  \frac{R}{z_\infty} \left[ \int_0^{\tilde{\ell}/2} d \tilde{x}
    \frac{\tilde{z}_\ast^2}{\tilde{z}(\tilde{x})^4} -
    \frac{1}{\tilde{\epsilon}} - \frac{\pi}{2 \tilde{\ell}} \frac{
      \Gamma(-1/4) \Gamma(3/4)}{\Gamma(1/4)^2} \right]\ ,
\end{equation} 
and with this subtraction we have chosen that the entanglement entropy
$\tilde{s}$ starts at zero in the infinite past.  We present some
results in figure \ref{fig:thermal_vary_v0}, showing the evolution for
a range of quench speeds, set by $\tilde{v}_0$.  We see that the
curves almost perfectly overlap when $\tilde{v}_0$ is small enough,
except in the region before saturation is reached, where the curve
with smaller $\tilde{v}_0$ appears to be sharper.  The invariance of
the early evolution suggests that regardless of how fast the quench is
done, the early evolution is dictated by a time constant that is only
dependent on the details of the plasma described by the dual 2+1
dimensional field theory.  This behavior was also observed in
refs.\cite{Hubeny:2007xt,AbajoArrastia:2010yt} for one dimension fewer.
Notice that for large $\tilde{v}_0$, the evolution of the entropy
appears to follow the evolution of the quench rather closely. For
short quenches the system relaxes to its new state more slowly.

\begin{figure}[h]
\begin{center}
   \includegraphics[width=3in]{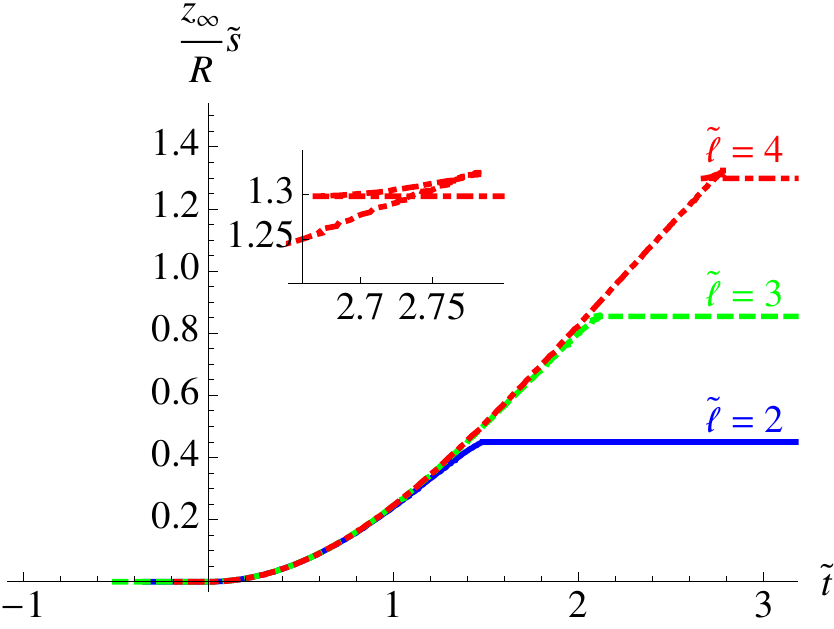} 
   \caption{\small Results for thermal quench for $\tilde{v}_0 = 0.01$.}  \label{fig:thermal_vary_ell}
   \end{center}
\end{figure}

\begin{figure}[h]
\begin{center}
\subfigure[Evolution of $\tilde{z}_\ast$]{\includegraphics[width=2.8in]{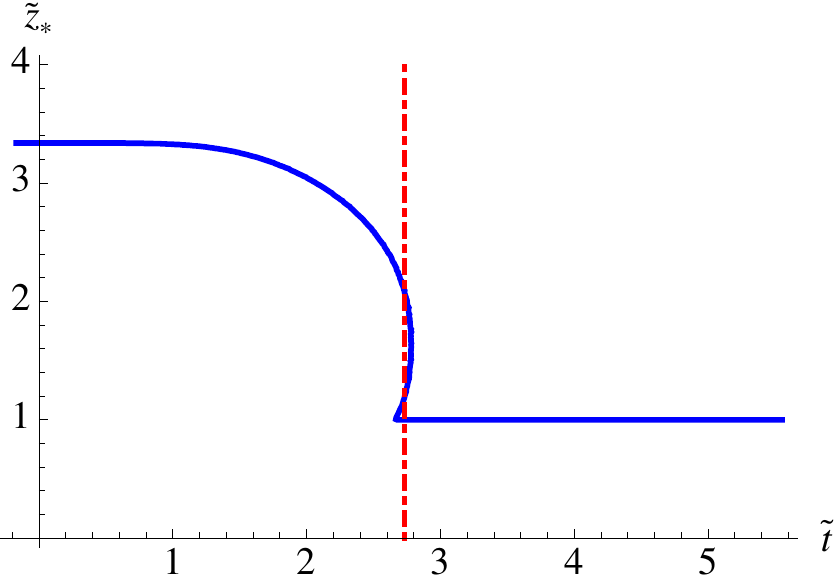} \label{fig:thermal_zstar}} \hspace{0.5cm}
\subfigure[Embeddings at the transition]{\includegraphics[width=2.8in]{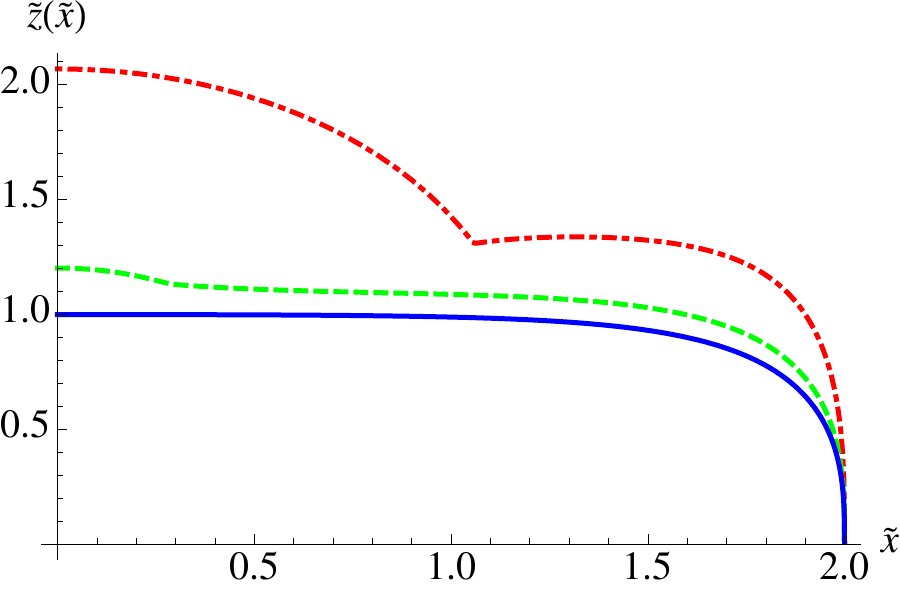} \label{fig:transition_embeddings}}
   \caption{\small Behavior of $\tilde{z}_\ast$ for $\tilde{\ell} = 4$ and  $\tilde{v}_0 = 0.01$.  In (a), the red dashed line is where the transition occurs and the apparent horizon is at $z=1$.  In (b), we display the two embeddings at the transition, where the red, dot--dashed curve corresponds to the embedding with $\tilde{z}_\ast = 2.0669$, the green, dashed curve corresponds to the embedding with $\tilde{z}_\ast = 1.2010$, and the blue, solid curve corresponds to the embedding with $\tilde{z}_\ast = 0.9985$}  \label{}
   \end{center}
\end{figure}

Exploring the behavior as we change the strip width $\tilde{\ell}$, we
note several things.  First, the early evolution appears to be
independent of (or only weakly dependent on) $\tilde{\ell}$. This can
be understood from the geometry of the strip.  As $\ell$ changes, the
area at the boundary of $\mathcal{A}$ does not change, so the initial
propagation of excitations from region $\mathcal{A}$ to region
$\mathcal{B}$, contributing to the entanglement, is not affected by
the strip width.  Second, we see the appearance of a new phenomenon,
the appearance of a swallow tail. See figure
\ref{fig:thermal_vary_ell}.  The swallow tail is a feature of our
numerical method for searching for minimal area solutions with a given
UV and IR boundary condition at a given time $\tilde{t}$. The method
reveals the presence of multiple (at times up to three) solutions to
the problem for a given time.  By studying figure
\ref{fig:thermal_zstar}, we see that the multiplicity corresponds to
the presence of solutions that have a $\tilde{z}_\ast$ below and above
the apparent horizon.  The solution is always the one with the lowest
area, and therefore the transition leads to a kink in the entropy
dependence on time.  Figure \ref{fig:transition_embeddings} explicitly shows the three embeddings at the transition; the transition is from an embedding with several sharp features to a smooth one.

This kink in the evolution of the entanglement entropy is a striking
new feature in time evolution. Notice that the branch of solutions
joined on to at late times are ones for which the turning point of the
minimal surfaces, $\tilde{z}_*$, is well above the final horizon,
$\tilde{z}_\infty$, giving an $\tilde{s}$ which is essentially the
saturated late time value. In other words, the entropy reaches
saturation and then abruptly flattens out for large enough strip size
$\tilde{\ell}$. For smaller strip size, the transition is also fast,
but not so fast as to have a discontinuity in the derivative.  Since the rate of approach to saturation is not strongly dependent on the region size $\tilde{\ell}$, while the saturation value itself depends on $\tilde{\ell}$, it is reasonable to expect that upon nearing saturation, the evolution turns over and flattens out toward the saturation value. In cases where the saturation value increases fast enough with $\tilde{\ell}$ (linearly with $\tilde{\ell}$ for the strip) while the rate of approach remains constant or weakly dependent on $\tilde{\ell}$, this turnover will be more pronounced for large enough $\tilde{\ell}$.  In these large $N$ systems we are studying, this is, we presume, the role of the kink, which as we have seen appears for large enough $\tilde{\ell}$.  We expect that the sharp transition would be softened by $1/N$ corrections upon leaving the
gravity approximation.

Our results suggest that the picture proposed in ref.~\cite{Calabrese:2005}, which arose in the 1+1 dimensional case, needs modification here.  In their picture, the entropy is discussed in terms of left and right moving pairs in the field theory. Saturation occurs once all pairs have had enough time to have traveled a distance of order the size of the interval. For our higher dimensional geometries we see saturation before all the analogous pairs are separated enough (here for the strip, for example, there are directions for which pairs can travel for extremely long times before leaving the strip). This suggests that the entanglement entropy is not built from simple pairwise contributions for the higher dimensional cases under consideration here. The bulk of the contributions from those longer-traveling quanta must have been already accounted for, and the system nears saturation without them having departed the starting region.

%

\begin{figure}[h] %
   \centering
   \includegraphics[width=3in]{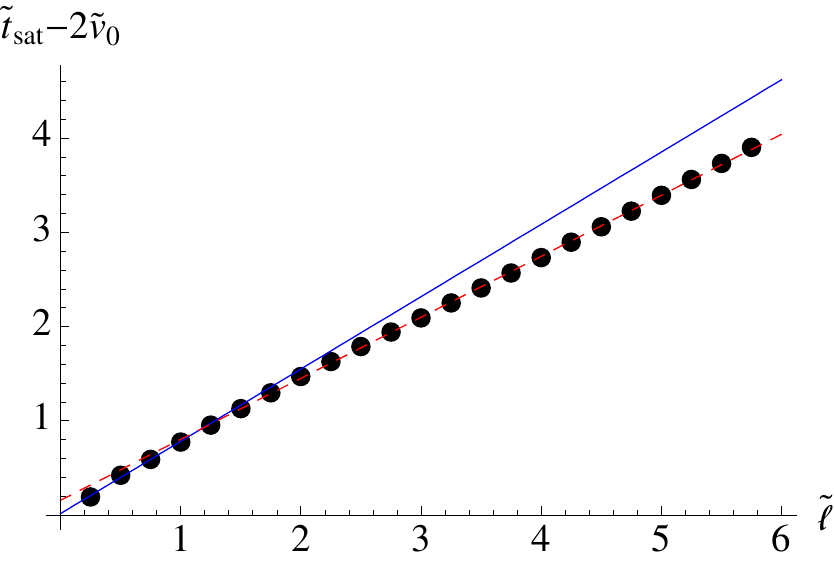} 
   \caption{\small The saturation time as a function of $\tilde{\ell}$ for thermal quench of the strip.  The solve of the best fit line for small $\tilde{\ell}$ depicted with the solid blue line is 0.77.  The slope of the best fit line for large $\tilde{\ell}$ depicted with the red dashed line is 0.65.}
   \label{fig:thermal_t_sat_vs_ell}
\end{figure}

Turning to the intermediate time region where the entanglement entropy
grows rapidly, we show in figure \ref{fig:thermal_t_sat_vs_ell}) the
behavior of the saturation time for various $\tilde{\ell}$.  Near
$\tilde{\ell} = 0$, the best fit line has slope 0.77, yet for much
larger $\tilde{\ell}$, the slope is closer to 0.65 (Note that there is
some flexibility in this number depending on how one defines the edges
of the growth period to extract the saturation time $t_{\rm
  sat}$). Interestingly, therefore, it seems that the saturation rate
for the thermal case grows faster for larger $\ell$.  This is
traceable to the early time behaviour of the curves. See figure
\ref{fig:thermal_vary_ell}.  For small $\tilde{\ell}$ (\emph{i.e.}$
\lesssim 2$), saturation is reached well before we enter the linear
phase , so we may expect differences in the saturation time for those
cases versus large $\tilde{\ell}$.

It is important to note that our saturation time is a departure from
the result of $\tilde{\ell}/2$ for the 1+1 CFT result of
ref.~\cite{Calabrese:2005}, reproduced in the holographic thermal
quench studies of refs.\cite{Hubeny:2007xt,AbajoArrastia:2010yt}. Note that here the
excitations do not necessarily travel a distance of order $ 2 \ell$
during this time.  Because of the geometry of the strip, some
excitations can travel much greater lengths.  Therefore, the overall
effective speed of saturation of $\tilde{s}$ should be expected to be
less than the speed of light, as we have seen in our results for the
strip.
%
\subsection{The Disc}
\begin{figure}[h]
\begin{center}
\subfigure[Thermal quench, with  $v_0 = 0.01$.]{\includegraphics[width=2.8in]{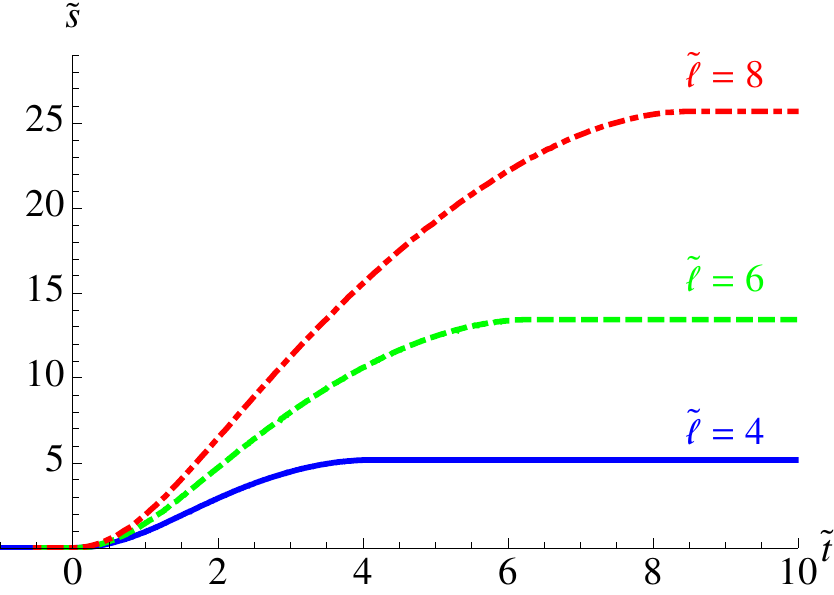} \label{fig:thermal_entropy_vary_ell_disk}} \hspace{0.5cm}
\subfigure[Saturation time for varying $\tilde{\ell}$, with $v_0 = 0.01$.]{\includegraphics[width=2.8in]{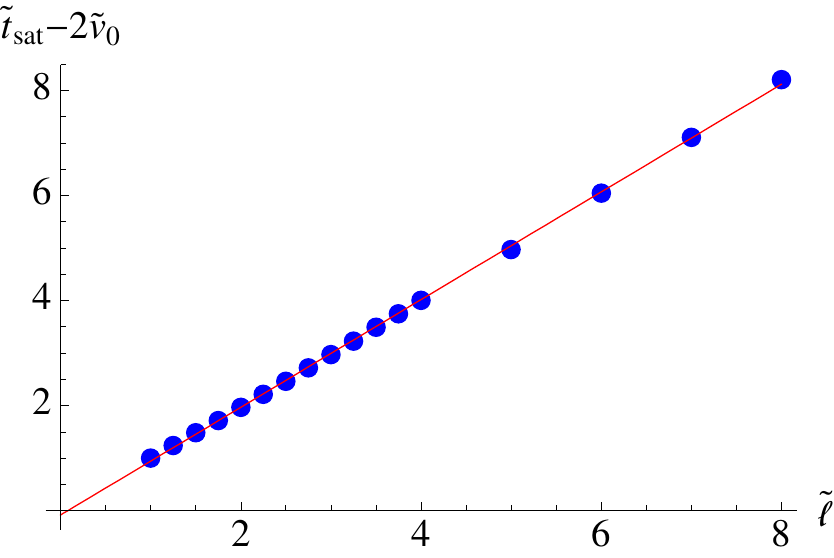} \label{fig:tsat_thermal_disk}}
   \caption{\small Results for thermal quench with a disk.  The slope of the saturation time is almost exactly unity.}  \label{fig:thermal_disk_results}
   \end{center}
\end{figure}

%
We can perform similar calculations for the disc.  The area of the two
dimensional surface is given by equation \reef{eqt:area_disc}, and we
define:
\begin{equation} \label{eqt:tilde_s_quench_disc}
\tilde{s} =  \int_0^{\tilde{\ell}} d \tilde{r} \frac{\tilde{r}}{\tilde{z}(\tilde{r})^2} \sqrt{ 1 - f(\tilde{v},\tilde{z}) \tilde{v}'(\tilde{r})^2 - 2 \tilde{v}'(\tilde{r}) \tilde{z}'(\tilde{r})} - \frac{\tilde{\ell}}{\tilde{\epsilon}} + 1
\end{equation}
We present some results in figure \ref{fig:thermal_disk_results}.
There are several differences from the strip case.  First, we see that
the early evolution strongly depends on $\tilde{\ell}$.  Once again,
this can be understood from the geometry of $\mathcal{A}$. The region
near the boundary of $\mathcal{A}$ has a size that grows with $\ell$
and so the early evolution of the entanglement should reflect this
dependence on geometry.  Another key difference from the strip case is
that no matter how large $\tilde{\ell}$ is, we do not see the kink in
the transition to saturation.  In line with our reasoning from the previous section, the saturation value of the entanglement entropy and the rate of approach to saturation increase in lock with $\tilde{\ell}$ (unlike for the strip where the rate was independent of $\tilde{\ell}$) allowing for a smooth evolution towards saturation.   Furthermore, we see that the saturation
time grows linearly (with a slope of unity) with $\tilde{\ell}$. This
fits with the fact that the disc's finite extent in all directions
means that there is a maximum travel time for any excitations as a
result of the quench to leave $\mathcal{A}$ and contribute to the
entanglement entropy. So in this way we recover that the rate of
saturation of the entanglement entropy is the speed of light, as was
observed in lower dimensions. In this sense, the disc is a more
natural generalization of the 1+1 dimensional case than the strip.

%
\section{The Electromagnetic Quench}
%
It is possible to cleanly study an entirely different type of quench
in this system. This non--thermal quench is achieved by turning on
electric and magnetic sources of the theory. We choose the system to
be entirely non--thermal by approaching the extremal ({\it i.e.,} zero
temperature) black hole solution at late times\footnote{Since the extremal solution is a highly degenerate point in the space of black hole solutions, it is clear that forming one by collapse is a delicate process.  However, it is possible, at least formally.  It turns out that the novel physics that we observe in this system does not appear to depend upon the details of the process of formation.}. The electric
components of the background at late times can be read off as
non--zero charge density and chemical potential in the dual 2+1
dimensional field theory\cite{Chamblin:1999tk,Chamblin:1999hg}, whilst
the magnetic components represent a background magnetic
field\cite{Hartnoll:2007ai}.  In fact, electric--magnetic duality
invariance of our system means that different choices of electric or
magnetic sources can be rotated into each other, and we will often
think of the system as undergoing a purely magnetic quench for
convenience, but admixtures of electric (chemical potential) and
magnetic (background field) quenches are captured by all of our
results.

The collapsing charged geometry has a metric similar to the extremal
AdS$_4$ metric from the section \ref{section:Dyonic}, but with the
function $f$ now taken to be:
\begin{equation}
f(v,z) = 1 - 4 \left( \frac{z}{z_0(v)} \right)^3 + 3 \left( \frac{z}{z_0(v)} \right)^4 \ ,
\end{equation}
 taking:
\begin{equation}
q_e(v)^2 + q_m(v)^2 = \frac{3 R^6}{z_0(v)^4}  \ , \quad m(v) = \frac{2 R^4}{z_0(v)^3} \ .
\end{equation}
We use the same quench profile for $z_0(v)$ as that given in equation
\reef{eqt:z0}.  Note that here $z_\infty$ sets the final strength of
the magnetic field (in the duality frame where we choose everything to
be magnetic).
%
\subsection{The Strip}
\begin{figure}[h]
\begin{center}
\subfigure[Entropy]{\includegraphics[width=3.0in]{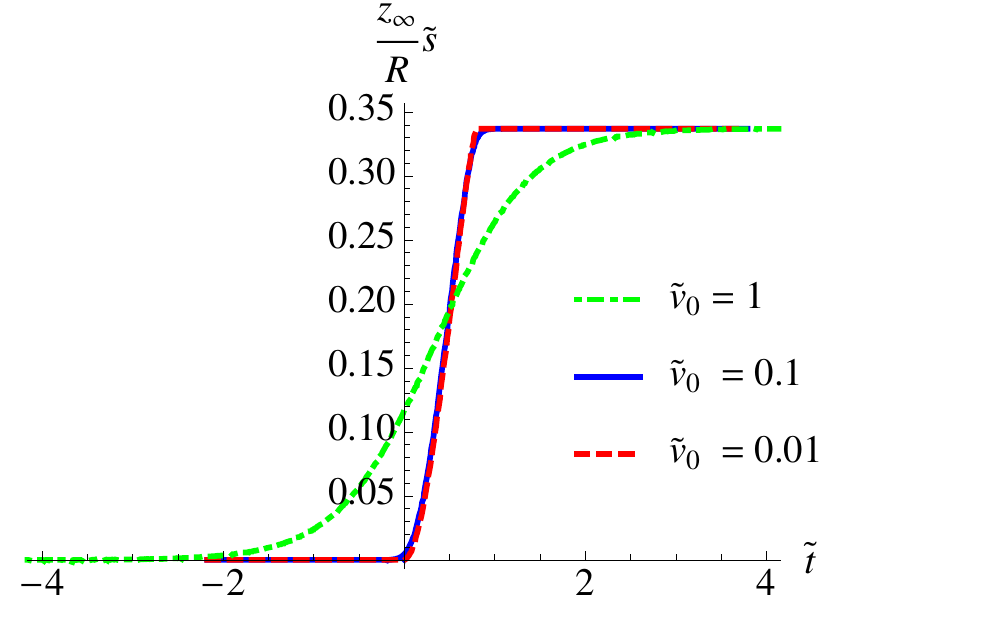}} \hspace{0.5cm}
\subfigure[Magnetic Field]{\includegraphics[width=3.0in]{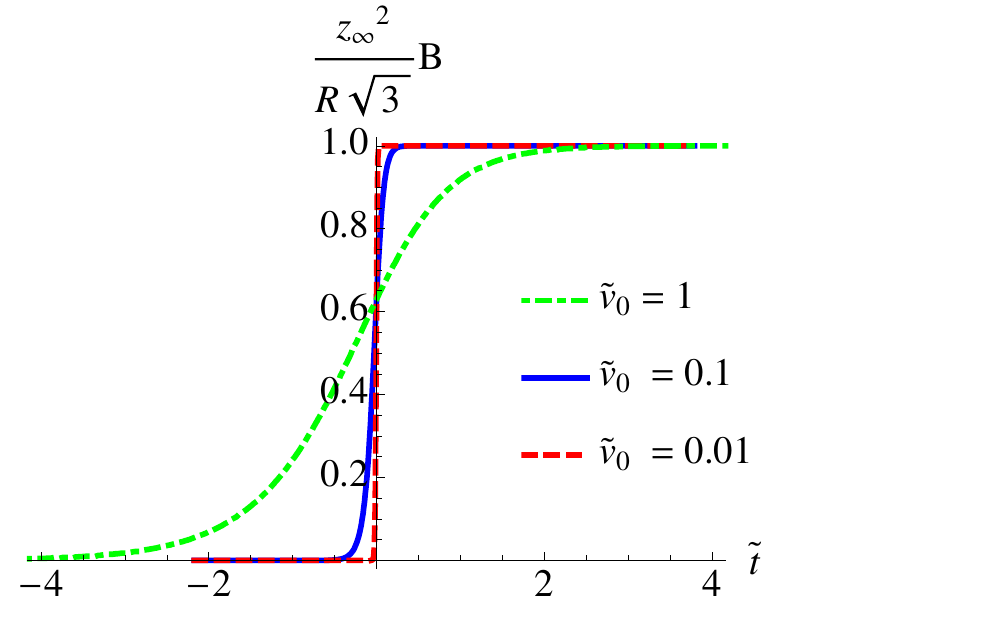}}
   \caption{\small The entropy and magnetic field for $\tilde{\ell} = 1$.}  \label{fig:v0_effect}
   \end{center}
\end{figure}

\begin{figure}[ht] %
   \centering
   \includegraphics[width=3in]{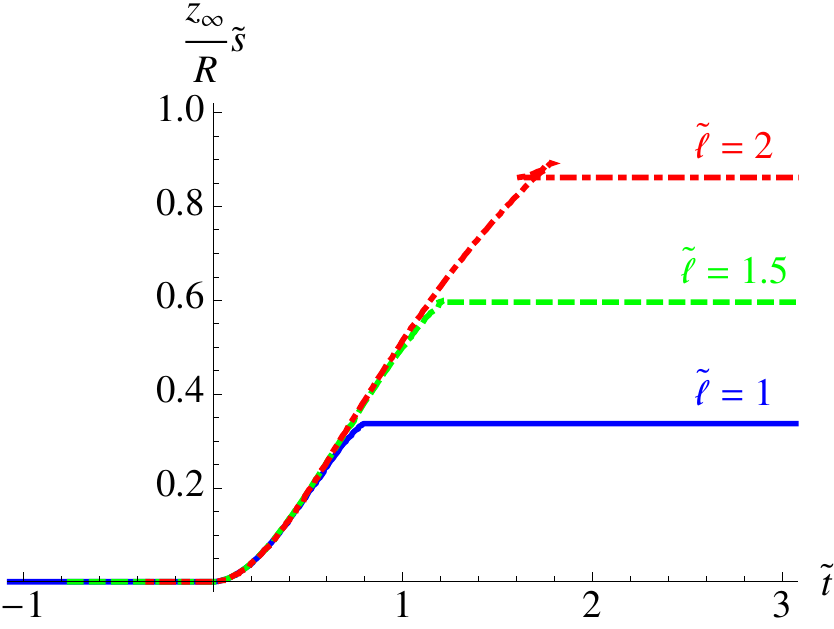} 
   \caption{\small  The entropy for various values of $\tilde{\ell}$ and $\tilde{v}_0 = 0.01$.}
   \label{fig:ell_effect}
\end{figure}

\begin{figure}[ht] %
   \centering
   \includegraphics[width=3in]{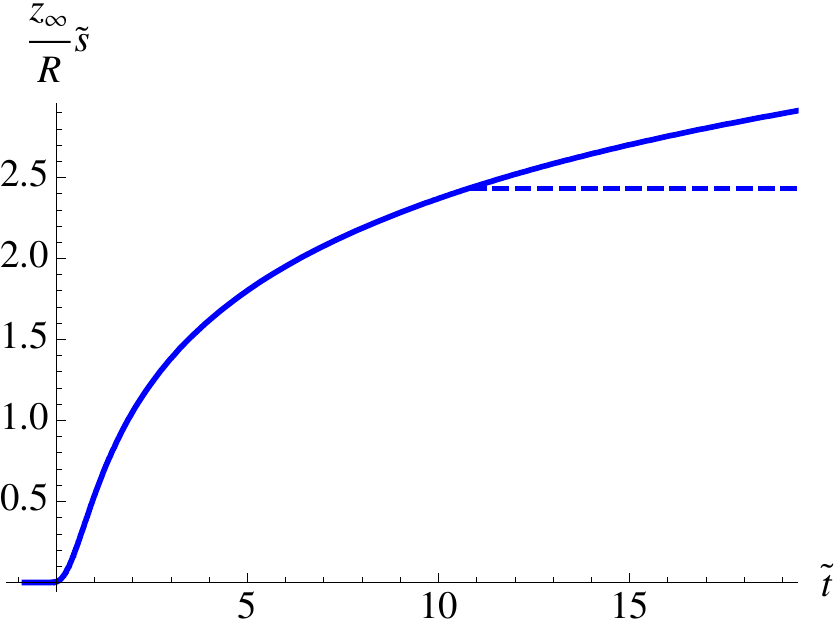} 
   \caption{\small The entropy for $\tilde{\ell}=5$ and $\tilde{v}_0 =
     0.1$.  The intermediate time behavior of the area is well
     approximated by $\ln \tilde{t}$.  The dashed line corresponds to
     the second embedding to which the system transitions, showing
     entanglement saturation.}
   \label{fig:ell=5}
\end{figure}

%
We proceed in a similar fashion as in the case of the thermal quench.
The area is given by equation~\reef{eqt:quench_area_strip}, and we
define the finite entanglement entropy $\tilde{s}$ by
equation~\reef{eqt:quench_s_strip}. Let us consider first the effect
of quench speed, set by $\tilde{v}_0$.  We present an example in
figure \ref{fig:v0_effect}. As with the thermal quench, it seems the
general behavior of the entanglement entropy for small enough
$\tilde{\ell}$ is qualitatively universal.
When $\tilde{v}_0$ is small enough, it appears that the rise time to
linear behavior does not depend on $\tilde{v}_0$, suggesting that there
is an intrinsic time scale for the dual plasma to react to changes in
the medium, as in the case of one dimension
fewer~\cite{AbajoArrastia:2010yt}.  It is interesting to note, however,
that the time scale in the current system appears to be much smaller
than that of ref.\cite{AbajoArrastia:2010yt}.  As in the thermal case, for large
$\tilde{v}_0$, the evolution of the entropy appears to follow the
evolution of the quench rather closely. 
We consider the effect of strip size $\tilde{\ell}$, presenting some
results in figure~\ref{fig:ell_effect}.  The early time evolution is
independent of $\tilde{\ell}$ just as we saw for the thermal case.
Furthermore, for large enough $\tilde{\ell}$, there is again the
appearance of a swallow tail, controlling the rapid flattening out to
saturation.  The resulting kink in the entanglement entropy has a similar origin to that discussed in section \ref{sec:thermal_strip}, although in this case the situation is more severe since the rate of approach is slowed down by the logarithmic behavior.
An increase in $\tilde{\ell}$ give an increases in size for the
swallow tail, and eventually the transition occurs even before the
beginning of the swallow tail region is ever reached. See an
example in figure \ref{fig:ell=5}.  

A new important feature is a departure from the linear behavior of the
evolution to a logarithmic growth.  This behavior is inherited from the logarithmic behaviour we saw in the static
extremal case in section \ref{section:Dyonic}, which has its origin in
the double zero at the horizon. On the dual field theory side, this
means significantly new behaviour for the excitations as they
propagate after the quench.  Presumably, the charged excitations are
slowed down by electromagnetic interactions, either from screening in
the presence of non--zero charge density or due to propagation in the
background magnetic field\footnote{Logarithm growth of entanglement entropy has been observed in (1+1)--dimensional systems quenched with an external field (see \emph{e.g.} ref.\cite{Eisler:2007}).  It would be interesting to explore the possibility of a connection to our results.}.

\begin{figure}[ht] %
   \centering
   \includegraphics[width=3in]{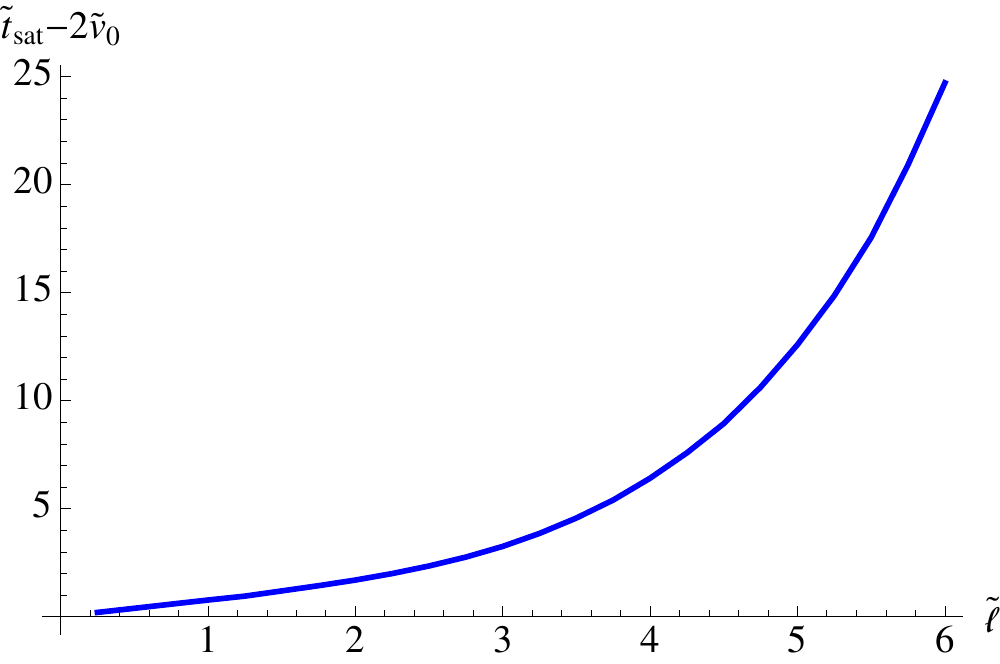} 
   \caption{\small The saturation time as a function of $\tilde{\ell}$ for $\tilde{v}_0 = 0.01$.  The slope near $\tilde{\ell} = 0$ is approximately 0.77.}
   \label{fig:t_sat_vs_ell}
\end{figure}

We can study how the saturation time $t_{\rm sat}$ depends on
$\tilde{\ell}$.  We present these results in
figure~\ref{fig:t_sat_vs_ell}.  We find that the slope of the line
near $\tilde{\ell} = 0$ is approximately 0.77 (note again that the
slope depends slightly on how one decides to calculate $t_{\rm sat}$
because of the roundedness of the solution before it saturates), which
matches our earlier result for the thermal quench.  Upon reaching the
point where the jump between embeddings starts to occur, the growth in
$\tilde{t}_{\rm sat}$ appears to be well fitted with an exponential
from that point onwards (to be expected from the logarithmic growth of
the area with time that we commented on earlier).
%
\subsection{The Disc}
\begin{figure}[ht]
\begin{center}
  \subfigure[Magnetic quench $v_0 =
  0.01$.]{\includegraphics[width=2.8in]{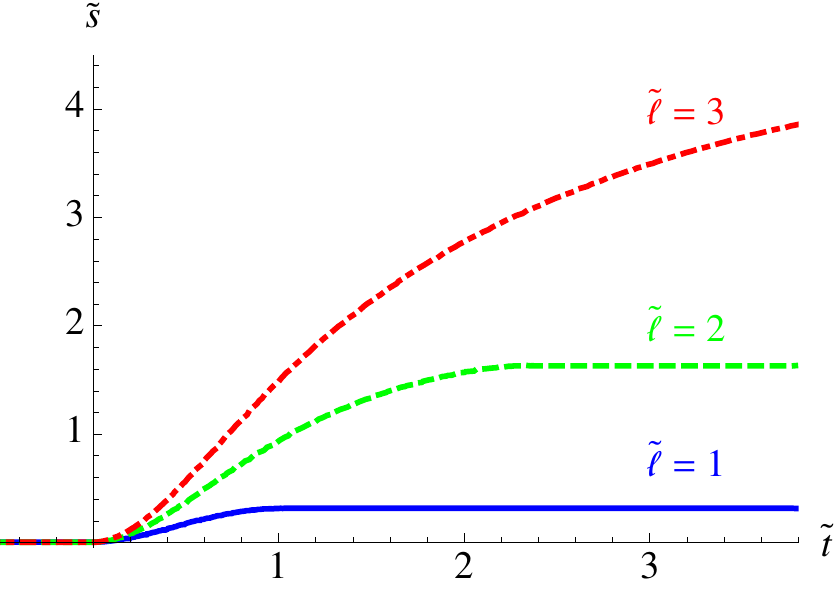} \label{fig:Entropy_vary_ell_v0=0_01_Disk}}
  \hspace{0.5cm} 
  \subfigure[Swallow tail, $\tilde{\ell} = 2.75$, $v_0= 0.01$.]{\includegraphics[width=2.8in]{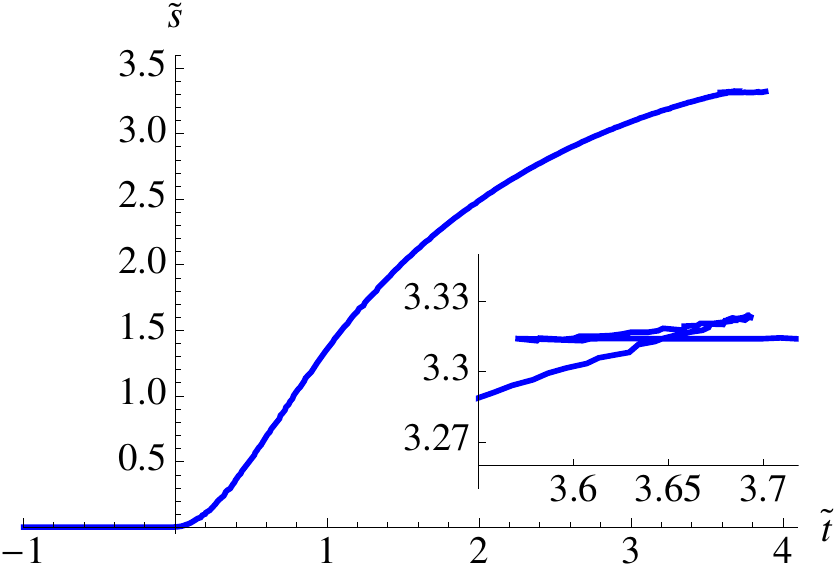} \label{fig:ell=2.75}}
  \hspace{0.5cm} \subfigure[Saturation time $v_0 =
  0.01$.]{\includegraphics[width=2.8in]{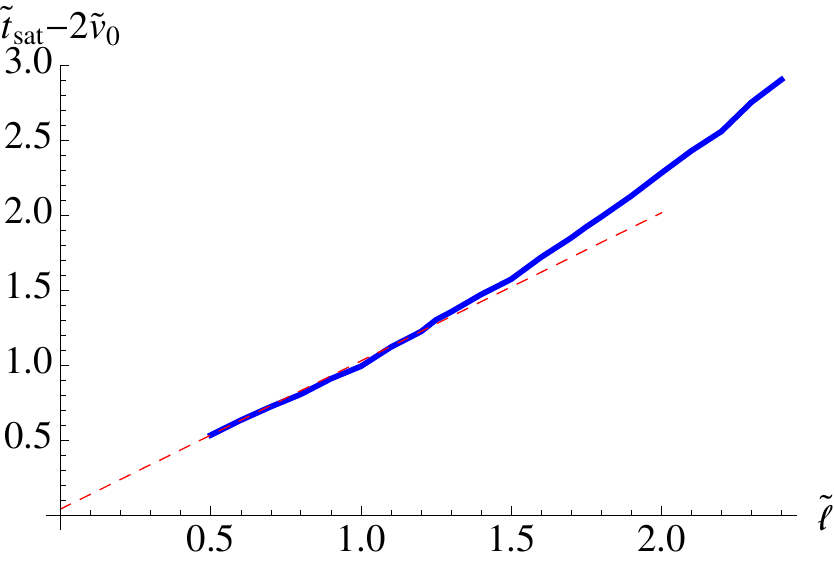} \label{fig:tsat_quench_disk}}
   \caption{\small Results for magnetic quench with a disc.  The slope of the saturation time for small $\tilde{\ell}$ is almost exactly unity.}  \label{fig:quench_disk_results}
   \end{center}
\end{figure}

%
We proceed in a similar fashion as in the case of the thermal quench.
The area is given by equation~\reef{eqt:area_disc}, and we define the
finite entanglement entropy $\tilde{s}$ by equation
\reef{eqt:tilde_s_quench_disc}.  We present some results in figure
\ref{fig:quench_disk_results}.  Again, the early evolution has
strong $\tilde{\ell}$ dependence in a manner attributable to the disc
geometry.  Also present is the logarithmic evolution, arising from the
near horizon region of the geometry.  Furthermore, for large enough
$\tilde{\ell}$, there is an appearance of the swallow tail. Recall
that this was absent for the thermal quench for the disc.    In line with the discussion in section \ref{sec:thermal_strip}, its appearance here is due to the slowing (as opposed to increasing in the thermal case) rate of approach to saturation with increasing  $\tilde{\ell}$ due to the logarithmic behavior.

\section{Conclusions}
We have studied two types of quench for the (2+1)--dimensional theory
dual to AdS$_4$, computing the evolution of the entanglement entropy
for an infinite strip region and a round disc region. Our quenches
were a thermal quench dual to a shell of uncharged matter (which forms
a Schwarzschild black hole) and a non--thermal quench, formed from a
shell of charged matter (which we tuned to form an extremal ({\it
  i.e.,} zero temperature) dyonic black hole.

We observed several interesting phenomena. The results for the disc,
for the thermal quench, provided the most direct analogue of earlier
observed results from (1+1)
dimensions\cite{Calabrese:2005,Hubeny:2007xt,AbajoArrastia:2010yt}: There was linear
growth in time of the entanglement entropy on its way to saturation, with a slope
consistent with the interpretation that the quanta responsible for
entanglement were propagating with the speed of light. The strip
deviated from this in two ways. The first, that the effective growth
rate (while still linear) was slower than the disc's was naturally
attributable to the semi--infinite geometry of the strip. The second
was more surprising. For large enough strip size there was a kink in
the entropy--time curve corresponding to a rapid turnover at
saturation.  The source of this on the geometry side is quite natural
(it is due to the availability of multiple branches of candidate
minimal surface solutions, forming a swallow tail shape --- there is a
jump from one branch to another at late enough times), but it would be
interesting to understand its origins on the field theory side.  The kink suggests a modification of the picture proposed in ref.~\cite{Calabrese:2005}, whereby not just entangled pairs contribute to the entanglement entropy.  (Note
that it is probable that such a kink gets softened beyond the
supergravity approximation.)

The kink is also present for both the disc and the strip for the
electromagnetic quench. One of the features of our novel quench is
that the entanglement entropy grows only logarithmically with time, a
feature that may be due (on the field theory side) to screening
effects for the charged quanta. It is interesting that kinks appear
when the effective propagation speed contributing to the entanglement
entropy growth is subluminal for both thermal and electromagnetic
cases, being absent for the thermal quench disc case. We do not know
if this is a coincidence, but are encouraged to pursue further study
of this issue.

In interpreting our results, we have assumed that the prescription to calculate the entanglement entropy is correct.  New features like the kink have been discussed from this point of view (see section \ref{sec:thermal_strip}).  We recognize that, until a better understanding of the prescription is obtained, it is possible that the kink may be interpreted as pathological, indicating a breakdown of the prescription to calculate the entanglement entropy for rapid quenches and large system size.

The logarithmic growth is again something that would be interesting to
understand further on the field theory side. On the gravity side it
comes from surfaces that probe near the extremal horizon, which has a
double zero in the metric function. A throat opens up at the horizon,
producing a smooth geometry that is AdS$_2\times \mathbb{R}^2$. It is
tempting to speculate that for our electromagnetic quench, the growth
of entropy in the intermediate region might be captured by an
effective lower dimensional theory captured by AdS$_2$ physics, as has
been shown to happen for aspects of the physics of studying Fermi
surfaces holographically in a related
context\cite{Faulkner:2009wj,Faulkner:2010tq}.

\section*{Acknowledgments}
TA would like to thank Stephan Haas for sparking our interest in this
work and for various useful discussions.  TA and CVJ would also like to thank Hubert Saleur, Silvano Garnerone, Rob Myers and Mukund Rangamani for useful discussions and comments as well as Esperanza Lopez for clarifications about aspects of
ref.\cite{AbajoArrastia:2010yt}.  This work was supported by the US Department of Energy.
%

\section*{Appendix: A Note  on Unitarity of Quench Evolution}
\begin{figure}[ht]
\begin{center}
\subfigure[Embedding 1]{\includegraphics[width=3.0in]{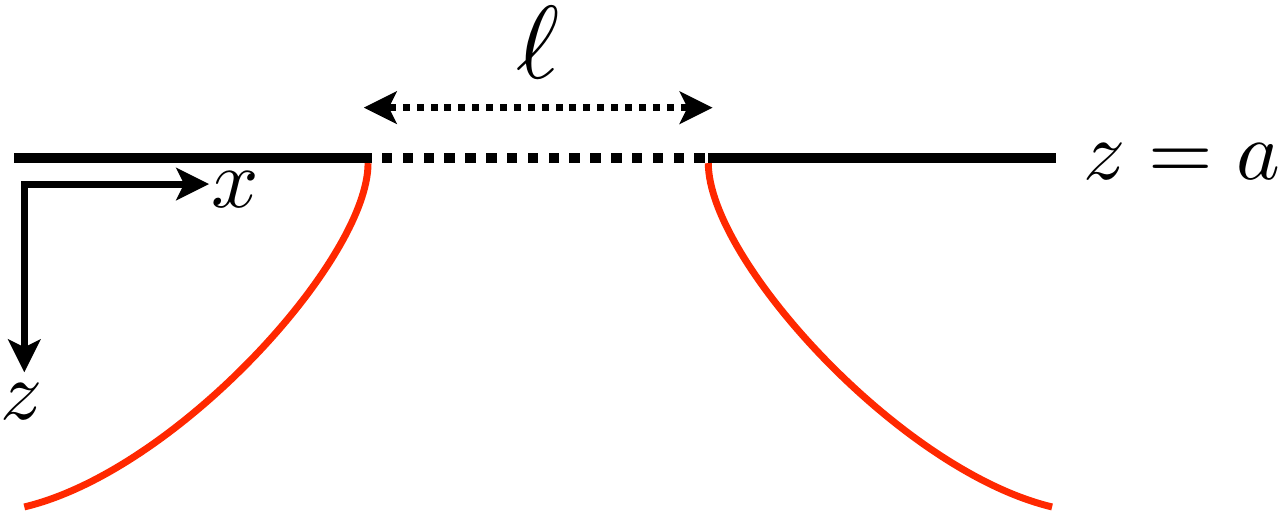} \label{fig:B_embedding1}} \hspace{0.5cm}
\subfigure[Embedding 2]{\includegraphics[width=3.0in]{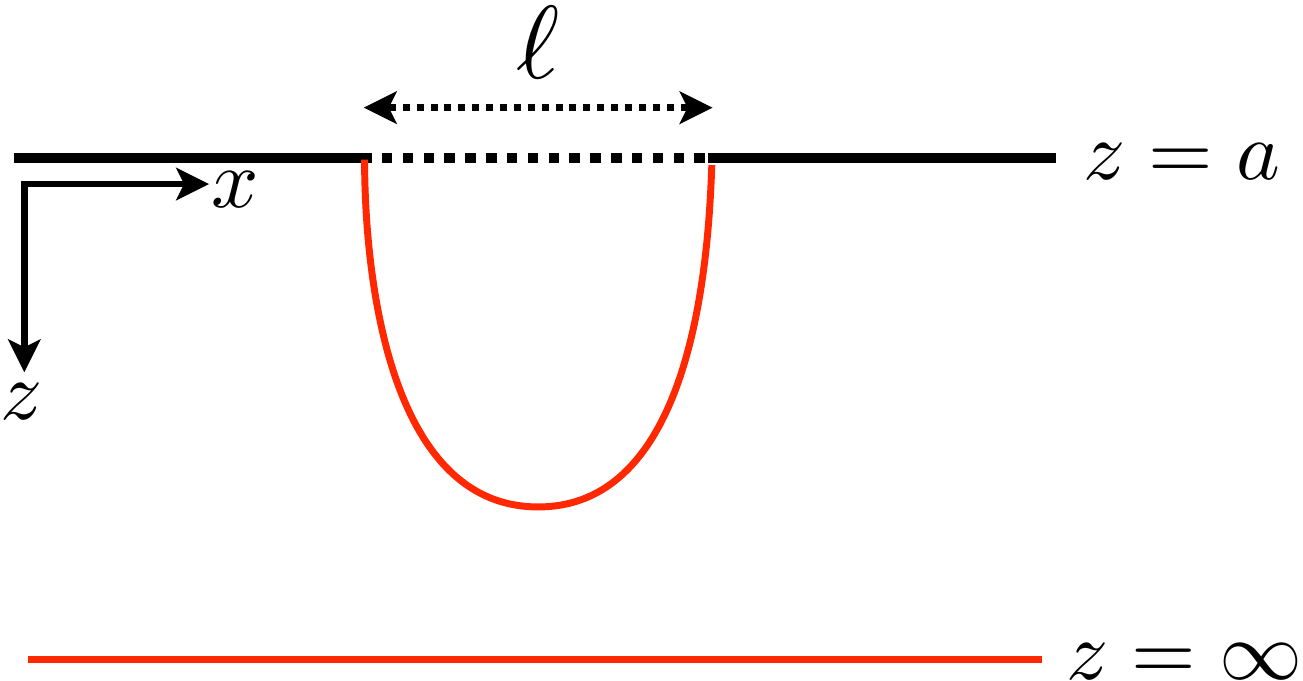}  \label{fig:B_embedding2}}
   \caption{\small Diagrams of the two types of embeddings to calculate entanglement entropy of $\mathcal{B}$.}  \label{fig:B_embeddings}
   \end{center}
\end{figure}

%
In order to show that the evolution of the quench and subsequent
relaxation process is unitary, we would like to show that throughout the
evolution the following condition is satisfied:
\begin{equation}
S_{\mathcal{A}} = S_{\mathcal{B}} \ .
\end{equation}
It is useful to start with some static examples before considering an
evolving background.  Let us begin with the case of pure AdS$_4$.  To
compute $S_{\mathcal{B}}$, there are two possible surfaces to consider
\cite{Headrick:2010zt}, which we depict in figure
\ref{fig:B_embeddings}.  The embedding in figure
\ref{fig:B_embedding1} can be shown to be divergent by first
considering it to be a sum of two finite intervals and then sending
the outer edges of the intervals to infinity.  The embedding in figure
\ref{fig:B_embedding2} has a piece that gives $S_\mathcal{A}$ and the
contribution from the piece at $z = \infty$ is zero.  Therefore, we
find that AdS$_4$ corresponds to a pure state.

Let us now consider a static Schwarzschild black hole.  Repeating the
calculation described above, we find that the second piece of figure
\ref{fig:B_embedding2} must lie not at $z = \infty$ but at $ z = z_0$,
the horizon.  Because of this fact, its contribution gives a non--zero
value; in fact it gives an entropy of:
\begin{equation}
S_{\mathcal{B}} = S_{\mathcal{A}} + \frac{1}{4 G_{\rm N}} \left( \frac{R^2}{z_0^2} \int dx dy  \right) = S_{\mathcal{A}} + \frac{\mathrm{Area(E.H.)}}{4 G_{\rm N}} \ ,
\end{equation}
where $\mathrm{Area}({\rm E.H.})$ is the area of the event horizon.
Therefore, we find that $S_{\mathcal{B}} \neq S_{\mathcal{A}} $, and
so the system is not in a pure state.  Therefore, any static solution
with an event horizon corresponds to a mixed state (see
e.g. ref.\cite{Myers:1997qi}).

So let us now consider our evolving metric.  If we consider a solution
that corresponds to the embedding of figure \ref{fig:B_embedding2}, a
solution to the equations of motion for the second disjoint piece is
to simply have:
\begin{equation}
z(x) = \infty \ , \quad v(x) = - \infty \ .
\end{equation}
We note that for $v(x) = - \infty$, our metric does not have a
singularity at $z(x) = \infty$, so there is no danger for the solution
to lie there.  In particular, this solution is (in a sense) simply the
solution that we would get for AdS$_4$ and gives zero contribution to
the entropy.  The first piece will give us the entropy of
$\mathcal{A}$.  Therefore, for the evolving case, we find that we
have:
\begin{equation}
S_{\mathcal{A}} = S_{\mathcal{B}} \ , \quad \forall t
\end{equation}
Note that this only works because we are assuming the second piece
remains static in the sense that it remains at $v = - \infty$ whereas
the other piece evolves.  Therefore, with this in mind, we find that
the evolution is unitary.
\providecommand{\href}[2]{#2}\begingroup\raggedright\endgroup

\end{document}